\documentclass[twocolumn]{aastex62}

\usepackage{graphicx}
\usepackage{txfonts}
\usepackage{verbatim}
\usepackage{natbib}
\usepackage[figuresright]{rotating}

\def\kms{km\,s$^{-1}$}

\def\msun{M$_{\odot}$}

\def\teff{T$_\mathrm{eff}$}

\def\l{$\lambda$}

\def\halpha{H~$\alpha$}

\def\hea{He\,{\sc i}}
\def\heb{He\,{\sc ii}}

\def\cc{C\,{\sc iii}}
\def\cd{C\,{\sc iv}}
\def\ce{C\,{\sc v}}

\def\nd{N\,{\sc iv}}

\def\od{O\,{\sc iv}}

\received{February 23, 2019}
\revised{-}
\accepted{May 25, 2019}
\submitjournal{ApJ}

\shorttitle{Atmosphere Analysis of VFTS 352}
\shortauthors{Abdul-Masih et al.}

\begin{document}

\title{Clues on the Origin and Evolution of Massive Contact Binaries: Atmosphere Analysis of VFTS 352}

\correspondingauthor{Michael Abdul-Masih}
\email{michael.abdulmasih@kuleuven.be}

\author{Michael Abdul-Masih}
\affil{Institute of Astrophysics, KULeuven, Celestijnenlaan 200 D, 3001 Leuven, Belgium}

\author{Hugues Sana}
\affil{Institute of Astrophysics, KULeuven, Celestijnenlaan 200 D, 3001 Leuven, Belgium}

\author{Jon Sundqvist}
\affil{Institute of Astrophysics, KULeuven, Celestijnenlaan 200 D, 3001 Leuven, Belgium}

\author{Laurent Mahy}
\affil{Institute of Astrophysics, KULeuven, Celestijnenlaan 200 D, 3001 Leuven, Belgium}

\author{Athira Menon}
\affil{Astronomical Institute Anton Pannekoek, Amsterdam University, Science Park 904, 1098~XH, Amsterdam, The Netherlands}

\author{Leonardo A. Almeida}
\affil{Departamento de F\'isica Te\'orica e Experimental, Universidade Federal do Rio Grande do Norte, CP 1641, Natal, RN, 59072-970, Brazil}

\author{Alex De Koter}
\affil{Institute of Astrophysics, KULeuven, Celestijnenlaan 200 D, 3001 Leuven, Belgium}
\affil{Astronomical Institute Anton Pannekoek, Amsterdam University, Science Park 904, 1098~XH, Amsterdam, The Netherlands}

\author{Selma E. de Mink}
\affil{Astronomical Institute Anton Pannekoek, Amsterdam University, Science Park 904, 1098~XH, Amsterdam, The Netherlands}


\author{Stephen Justham}
\affil{Astronomical Institute Anton Pannekoek, Amsterdam University, Science Park 904, 1098~XH, Amsterdam, The Netherlands}
\affil{School of Astronomy \& Space Science, University of the Chinese Academy of Sciences, Beijing 100012, China}

\author{Norbert Langer}
\affil{Argelander-Institut f\"ur Astronomie, Universit\"at Bonn, Auf dem H\"ugel 71, 53121 Bonn, Germany}

\author{Joachim Puls}
\affil{LMU M\"unchen, Universit\"atssternwarte, Scheinerstr.~1, 81679 M\"unchen, Germany}

\author{Tomer Shenar}
\affil{Institute of Astrophysics, KULeuven, Celestijnenlaan 200 D, 3001 Leuven, Belgium}

\author{Frank Tramper}
\affil{Institute for Astronomy, Astrophysics, Space Applications \& Remote Sensing, National Observatory of Athens, P. Penteli, 15236 Athens, Greece}



\begin{abstract}

The massive O4.5 V + O5.5 V binary VFTS 352 in the Tarantula nebula is one of the shortest-period and most massive overcontact binaries known. Recent theoretical studies indicate that some of these systems could ultimately lead to the formation of gravitational waves via black hole binary mergers through the chemically homogeneous evolution pathway. By analyzing ultraviolet-optical phase-resolved spectroscopic data, we aim to constrain atmospheric and wind properties that could be later used to confront theoretical predictions from binary evolution. In particular, surface abundances are powerful diagnostics of the evolutionary status, mass transfer and the internal mixing processes. From a set of 32 VLT/FLAMES visual and 8 HST/COS ultraviolet spectra, we used spectral disentangling to separate the primary and secondary components. Using a genetic algorithm wrapped around the NLTE model atmosphere and spectral synthesis code \textsc{fastwind}, we perform an 11-parameter optimization to derive the atmospheric and wind parameters of both components, including the surface abundances of He, C, N, O and Si. We find that both components are hotter than expected compared to single-star evolutionary models indicating that additional mixing processes may be at play. However the derived chemical abundances do not show significant indications of mixing when adopting baseline values typical for the system environment.  


\end{abstract}



\section{Introduction} \label{sec:intro}

Massive stars are some of the most energetic cosmic engines driving the evolution of our universe.  Because most of these stars are found in relatively close multiple systems \citep[see e.g.,][for a review]{Sana2011}, binary interactions play an important role in their evolution. Based on galactic studies, about 40\% of all stars born as O-type are expected to interact with a companion before leaving the main sequence \citep{Sana2012}. Ignoring particular evolutionary pathways such as those involving chemically homogeneous evolution or magnetic fields (though see below), more than half of these interactions (about a quarter of all O-type stars) will lead to an overcontact phase and subsequent coalescence.  Despite affecting such a large percentage of O-type stars, the overcontact phase is one of the least understood phases of massive star evolution. This is due to the interplay of several complex physical processes occurring simultaneously during this phase, including, but not limited to, mass exchange, internal mixing, intense radiation fields, and potential internal structure adjustments \citep{Pols1994, Wellstein2001, deMink2007}.  In addition to the complex physics, our limited understanding also stems from a lack of observational constraints due to the short lived overcontact phase, which is expected to evolve on thermal timescales.  Currently there are only eight O-type overcontact systems known \citep{Almeida2015, Popper1978, Lorenzo2017, Howarth2015, Hilditch2005, Penny2008}

Aside from binary interaction, stellar rotation can also have a dramatic influence on the evolution of massive stars. As a rapidly rotating massive star evolves, the core will shrink and the envelope will expand, leading to differential rotation \citep{vonZeipel1924, Eddington1925, Eddington1926, Vogt1925}.  Differential rotation can in turn cause shear instabilities that produce significant mixing \citep{Zahn1974}.  Additionally, rapidly rotating stars experience surface deformations which can lead to Eddington-Sweet circulations due to the thermal imbalance between the poles and the equatorial regions.  These large meridional currents and the shear instabilities allow fresh material to be mixed into the core, extending the lifetime of the star. They also allow for nuclear burning products to be transported to the surface, leading to detectable abundance anomalies \citep{Maeder2000a, Maeder2000b, Brott2011a}.  Additionally, evolutionary models that include rotation show that with additional mixing, the effective temperature of a star increases \citep{Yoon2005, Yoon2006}.  
		
The efficiency of the rotational mixing  depends on the star's rotation rate,  mass and initial metallicity. If the mixing is efficient enough, it can prevent the build up of a chemical gradient at the interface between the core and envelope, allowing the star material to be fully mixed. Such a chemically homogeneous evolution (CHE) induced by rotational mixing was first proposed by \citet{Maeder1987}. In the CHE regime, a star will evolve up and to the left on an Hertzsprung-Russell diagram (HRD), at first moving up along the zero-age main sequence (ZAMS), then to the left towards hotter temperatures and at roughly constant bolometric luminosities.  If the mixing is not efficient enough, a star can also undergo  incomplete chemically homogeneous evolution (incomplete CHE) where the star will initially evolve up and to the left but will eventually turn back and evolve more typically.  Since CHE is dependent on the rotation rate, lower metallicities are favored for this pathway in the case of single stars \citep{Yoon2005, deMink2009, Brott2011b, Kohler2015, Szecsi2015}.  Massive stars have stellar winds driven by ultraviolet (UV) metallic lines. Therefore, low metallicity means weaker winds and thus less angular momentum lost from the winds.
	
While initially introduced for fast rotating single stars, CHE can also occur in massive binary systems.  In very close binary systems, the component stars can be tidally locked and their surface rotation rates determined by the orbital period and the stars radii. As first suggested by \citet{deMink2009}, CHE could delay or prevent coalescence  as the stars remain compact as they evolve, and might thus stay within their Roche lobes \citep[see also][]{Almeida2015}. Just before the first direct detection of gravitational waves \citep{Abbott2016}, CHE was proposed to be a possible pathway to the formation of black hole binary systems close enough to merge in a Hubble time \citep{deMink2016, Mandel2016, Marchant2016}. Two massive binary systems to date, VFTS 352 \citep{Almeida2015} and HD 5980 \citep{Koenigsberger2014},  show indications of CHE. This paper focuses on VFTS~352 in the Tarantula Nebula (a.k.a.\ 30 Doradus or 30 Dor) in the Large Magellanic Cloud (LMC).

VFTS 352 is an eclipsing double-lined O-type spectroscopic binary system discovered in the course of the VLT-Flames Tarantula Survey \citep[VFTS,][]{Evans2011, Sana2013}. With spectral types of O4.5 V(n)((fc))z + O5.5 V(n)((fc))z)  \citep{Walborn2014}, both stars are rather rapid rotators with projected rotation rates of approx. 290~\kms~\citep{Oscar2015}. Further spectroscopic monitoring as part of the Tarantula Massive Binary Monitoring \citep{Almeida2017} allowed for a first characterization of the orbit, yielding a short orbital period of 1.124 days. Using a Wilson Devinney code to fit the light and radial velocity curves,  \citet{Almeida2015} deduced stellar masses of about 29\msun\ and showed that the system was in deep contact. With effective temperatures of $42540 \pm 280$~K and $41120 \pm 290$~K, \citet{Almeida2015} found that the component stars were too hot for their dynamical masses, suggesting the presence of strong mixing and making  VFTS 352 a good candidate for CHE.  
		
In this paper, we investigate the evolutionary status of VFTS 352 in greater detail by fitting synthetic spectra to optical and UV data in order to constrain the physical and atmospheric properties of the system components, including the surface abundances of helium, carbon, nitrogen and oxygen that  are powerful tracers of mixing.  This paper is organized as follows.  Section \ref{sect: obs} describes our data set and provides the journal of the spectroscopic observations.  Section \ref{sect: disentangling} details the disentangling methods including the removal of interstellar and nebular lines.  Section \ref{sect: atm} describes the atmosphere analysis techniques and fitting methods, the results of which are discussed in Sect.\ \ref{sect: results}. Finally, Sect.\ \ref{sect: evol} addresses the evolutionary status of the system and  Sect.\ \ref{sect: ccl}  summarizes our conclusions.

\section{Observations} \label{sect: obs}

   \subsection{Optical Spectroscopy}
   
      The optical data consist of 32 well phase covered spectra collected over an 18-month period as part of the Tarantula Massive Binary Monitoring \citep[TMBM, PI: Sana H., ESO programs: 090.D-0323 and 092.D-0136][]{Almeida2017}, using the FLAMES-GIRAFFE multi-object spectrograph on the ESO Very Large Telescope (VLT).  The observations were performed using the LR02 setup, which provides continuous wavelength coverage from 3950\AA\ to 4550\AA\ with a spectral resolving power of 6,400.  Each of the 32 epochs consists of three back-to-back exposures of 900 seconds. These data were reduced using the ESO CPL GIRAFFE pipeline v.2.12.1, which includes bias subtraction, flat-field correction, spectral extraction and wavelength calibration.  Sky corrections and normalization procedures were applied following the description in \citet{Evans2011} and \citet{Sana2013}, respectively. The resulting spectra have signal-to-noise ratios within the same order of magnitude ($\sim 150$). The journal of the TMBM observations is given in \citet{Almeida2017}.  Figure \ref{raw_spectra} (bottom panel) shows one epoch of the optical spectra at a phase of $\phi \approx 0.75$.
      
   \subsection{UV Spectroscopy}
\begin{figure*}
   \centering
   \includegraphics[width=1\linewidth]{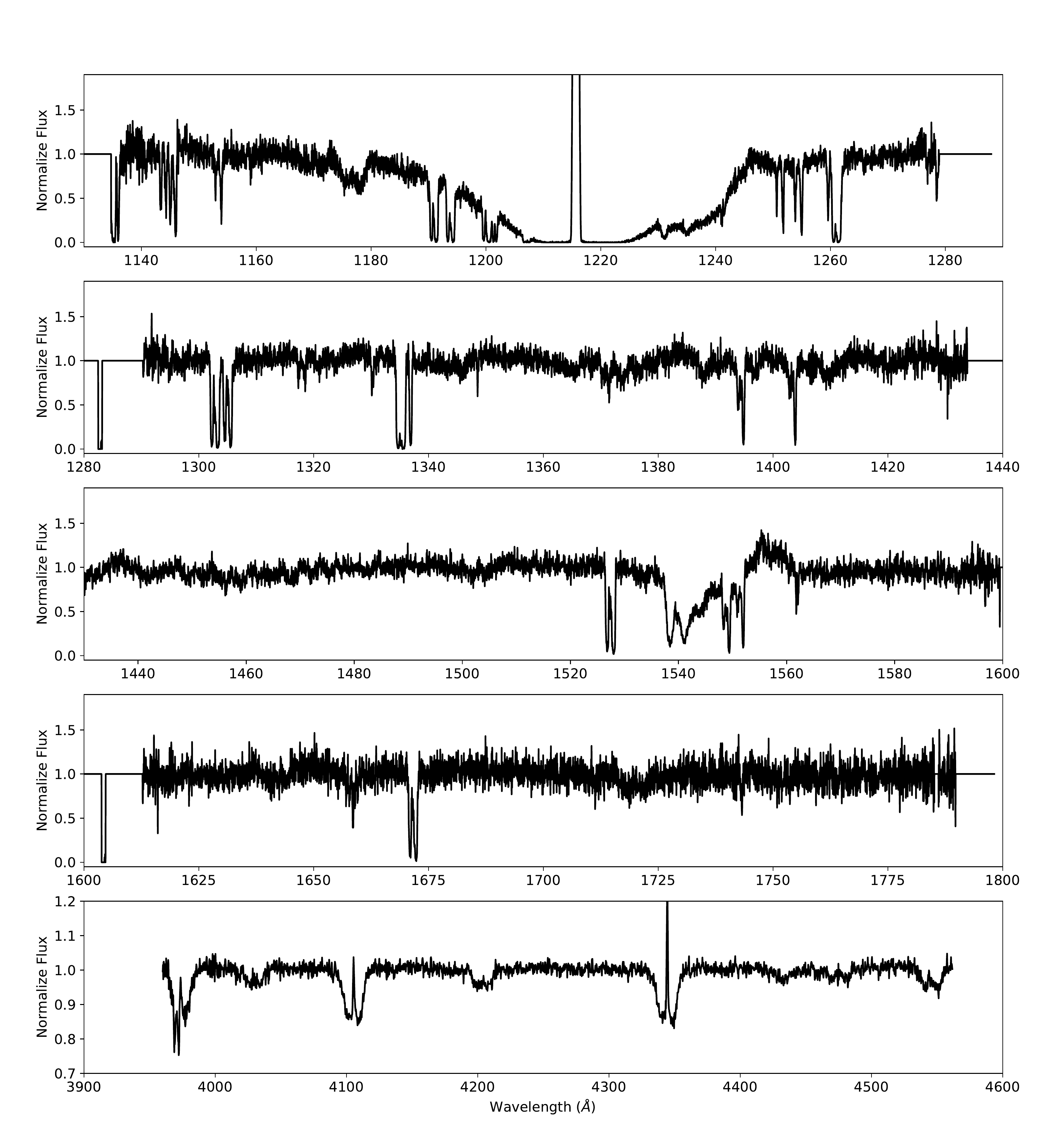}
      \caption{Sample of normalized, pre-disentangled UV and optical spectra of VFTS 352 at a phase of approximately 0.75.
              }
         \label{raw_spectra}
   \end{figure*}

        UV spectra were obtained with the Far-UV G130M and G160M gratings of the Cosmic Origins Spectrograph (COS) on the Hubble Space Telescope (HST) under the auspices of program GO 13806 (PI Sana). In total, eight spectra covering the entire orbit by bins of 0.125 in phase were obtained over a 10-day period.  The instrumental setup provides nearly continuous wavelength coverage from approximately 1130\AA\ to 1790\AA\ with a central wavelength of 1291\AA\ and resolving power of 18,000 for the G130M grating and a central wavelength of 1611\AA\ and a resolving power of 19,000 for the G160M grating.  The observations were completed in eight visits of one orbit each. Each visit consisted of $2 \times 450$~s exposures with the G130M grating followed by $2 \times  750$~s exposures with the G160M grating for a total exposure time ($t_\mathrm{exp}$) of 900 seconds and 1500 seconds, respectively.  The observations were collected using the TIME-TAG mode with the Primary Science Aperture in lifetime position 3.  For visits 1, 3, 6 and 8, FP positions 1 and 3 were used while for visits 2, 4, 5 and 7, FP positions 2 and 4 were used. These data were uniformly processed in the standard fashion by CALCOS 3.0 (2014-10-30).  Processing steps included correcting the photon-event table for dead-time and position effects caused by drifts in the detector electronics, geometric distortion, and the orbital Doppler shift of the observatory; binning the time-tag data; wavelength calibration based on the Pt-Ne spectra simultaneously acquired during each visit; extraction and photometric calibration of the 1-D spectra; and “shift and add” combination of spectra obtained at different FP-POS positions for both grating settings. The resulting spectra all have similar signal-to-noise ratios on the order of $\sim 11.5$. An overview of the UV observations is given in Table \ref{table:obs}.  Figure \ref{raw_spectra} (top four panels) shows one epoch of the UV spectra at a phase of $\phi \approx 0.75$.

\begin{table}
\caption{Journal of HST/COS UV Observations}
\centering 
\begin{tabular}{ccccc}
\hline\hline
Visit & Grating & HJD at & $\phi$ \tablenotemark{1} & $t_\mathrm{exp}$ \\
 & &  mid observation & & (s) \\
\hline
1 & G130M & 2457158.08817 & 0.477 & 900 \\
  & G160M & 2457158.12413 & 0.509 & 1500 \\
2 & G130M & 2457158.23676 & 0.609 & 900 \\
  & G160M & 2457158.26459 & 0.634 & 1500 \\
3 & G130M & 2457158.38582 & 0.742 & 900 \\
  & G160M & 2457158.40381 & 0.758 & 1500 \\
4 & G130M & 2457165.26816 & 0.864 & 900 \\
  & G160M & 2457165.28615 & 0.880 & 1500 \\
5 & G130M & 2457156.40494 & 0.980 & 900 \\
  & G160M & 2457156.42294 & 0.996 & 1500 \\
6 & G130M & 2457155.41174 & 0.096 & 900 \\
  & G160M & 2457155.42974 & 0.112 & 1500 \\
7 & G130M & 2457160.05892 & 0.230 & 900 \\
  & G160M & 2457160.08914 & 0.257 & 1500 \\
8 & G130M & 2457160.19075 & 0.347 & 900 \\
  & G160M & 2457160.22104 & 0.374 & 1500 \\
 \hline
\end{tabular}

\tablenotetext{1}{Computed using the ephemerides of \citet{Almeida2015} (see Table \ref{table:orb_sol})}
\label{table:obs}
\end{table}

   \subsection{Optical Photometry}
   
      Over 10 years of OGLE I-band and V-band photometry have been obtained for VFTS 352 as part of the OGLE-III and OGLE-IV projects \citep{Udalski2008}.  The full photometric data set is discussed in \citet{Almeida2015}, however for this study, we use exclusively the I-band photometry as there are many more measurements available in the I-band than V-band (760 versus 90 respectively).  As discussed in \citet{Almeida2015}, the light curve shows ellipsoidal variations and the subsequent analysis performed shows that both components are filling their Roche lobes and the system is in an over-contact configuration. In this paper, we exclusively use the photometry to estimate the dilution factor of spectral lines as a function of the orbital phase to inform the spectral disentangling described below. 
      
\section{Spectral Disentangling } \label{sect: disentangling}

   An important difficulty that arises when attempting to fit synthetic spectra to those observed from double lined spectroscopic binaries (SB2) systems is the fact that observations show a combination of the two component signals. With multiple well phase covered observational epochs, it is possible to disentangle the signals from the two components and obtain separate spectra for each. This can be accomplished with a variety of spectral disentangling methods \citep[for a review, see][]{Pavlovski2010}. While there are several codes well suited for this problem \citep[see for example][]{Simon1994, Hadrava1995, Hadrava2009, Skoda2011}, we choose the FDBinary code which  uses Fourier techniques in combination with the orbital parameters and the radial velocities of the components to disentangle the spectra \citep{Ilijic2004}.  FDBinary allows for three component disentangling as well as light ratio variations by phase, which, as we discuss below, will be  essential for our analysis.

   The strong nebulosity in the Tarantula nebula causes nebular emission lines to be present in the observed spectra of VFTS 352.  These cannot be properly subtracted because the FLAMES sky fibers do not sample their spatial variations with sufficient accuracy.  The quality of the corrections further varies for one epoch to another as a result of small pointing inaccuracies, variable seeing and atmospheric diffraction \citep{Evans2011}. This presents a problem when disentangling as residuals from the nebular correction contaminate the  profiles of lines of interest in the optical part of the spectrum.  If disentangling is performed on  contaminated lines, the nebular lines will artificially narrow the final disentangled line profiles of the two component stars. This can affect the stellar parameters derived from the disentangled spectra, specifically temperature and surface gravity.  For this reason, it is important to properly remove the contribution of the nebular emission lines before performing a two component disentangling.
   
   As mentioned above, the nebular contamination is not constant over the 32 spectra. The sky subtraction procedure described in \citet{Evans2011} removes a large portion of the nebular contamination but fails to remove all of it.  The final spectra still contain nebular residuals that must be accounted for. To do so, our disentangling procedure consists of three major steps: determination of average nebular line profiles, scaling and removal of the average nebular line  profiles, and finally a two component disentangling. We describe these three steps below. 
   
   In the following, we adopt the circular orbital solution of \citet{Almeida2015} for the disentangling (for convenience, the orbital elements are included here in Table \ref{table:orb_sol}).  This orbital solution was obtained through a full Wilson Devinney fit of both the V and I band OGLE photometry as well as the 32 radial velocity points obtained from the TMBM program.  More recently, \citet{Almeida2017} also derived an orbital solution of this system based on 37 radial velocity points, however this did not take the photometric information into account, and for this reason, we choose to use the solution of \citet{Almeida2015}. It is important to note that in both \citet{Almeida2015, Almeida2017} the radial velocities were determined by fitting Gaussian profiles to each component, with the constraint that the Gaussians for each component were identical for all phases.  For an overcontact system, this method does not take into account the light ratio variations throughout the orbit, which can lead to minor uncertainties in the final radial velocities.
   
\begin{table}
\caption{Circular orbital solution from \citep{Almeida2015}}
\centering 
\begin{tabular}{cc}
\hline\hline
Parameter & Value \\
\hline
$P_\mathrm{orb}$ \ (day) & $1.1241452 \pm 0.0000004$ \\
$T_\mathrm{0}$ \ (HJD)   & $2455261.119 \pm 0.003$ \\
$K_\mathrm{1}$ (\kms )   & $327.8 \pm 2.0$ \\
$K_\mathrm{2}$ (\kms )   & $324.5 \pm 2.0$ \\
 \hline
\end{tabular}

\label{table:orb_sol}
\end{table}
   
   \subsection{Nebular Line Profile Determination}
      
\begin{figure}
   \centering
   \includegraphics[width=1\linewidth]{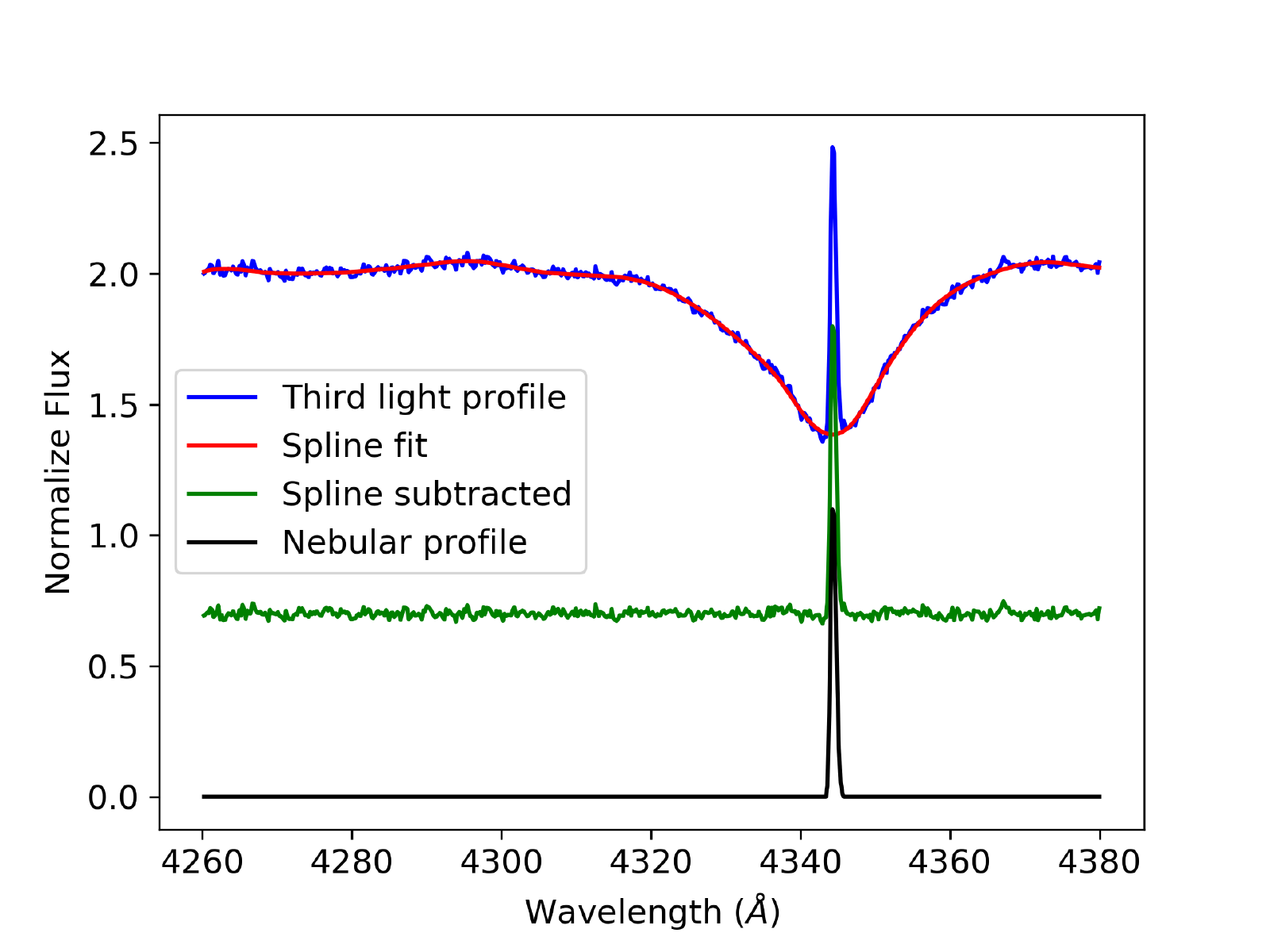}
      \caption{The profile of the third light component following a three component disentangling is plotted in the upper part of the top of the plot, with the spline fit to the profile (after cutting the nebular line) overplotted.  The spline-subtracted profile is shown in the middle.  The bottom spectrum shows the final nebular line profile after continuum noise has been clipped. The  spectra have been vertically shifted for clarity.
              }
         \label{spline_fit}
   \end{figure}

      The first step of the disentangling procedure involves the determination of the average nebular line profiles.  This can be accomplished using a three component disentangling where the third component is given a constant, null Doppler velocity.  This assigns all signals without radial velocity variations to the third component. Since VFTS 352 is an overcontact system, there are regions of the envelope which have radial velocity variations very close to zero, so some of this signal will be included in the disentangled third component spectra. Because we are only concerned with the shape of the nebular line profiles, we remove the signal from other sources.  As illustrated in Fig.~\ref{spline_fit}, this is done by selecting a small portion of the third light spectra around each of the nebular lines and then fitting a spline to the rest of the spectra.  We then subtract the spline from the original data, which yields the profiles of the nebular lines. Finally, we remove the noise for the regions outside of the nebular line profiles using the same cuts that were used before the spline was fit.  This gives the final nebular line profiles.
	      
   \begin{figure*}
   \centering
   \includegraphics[width=1\linewidth]{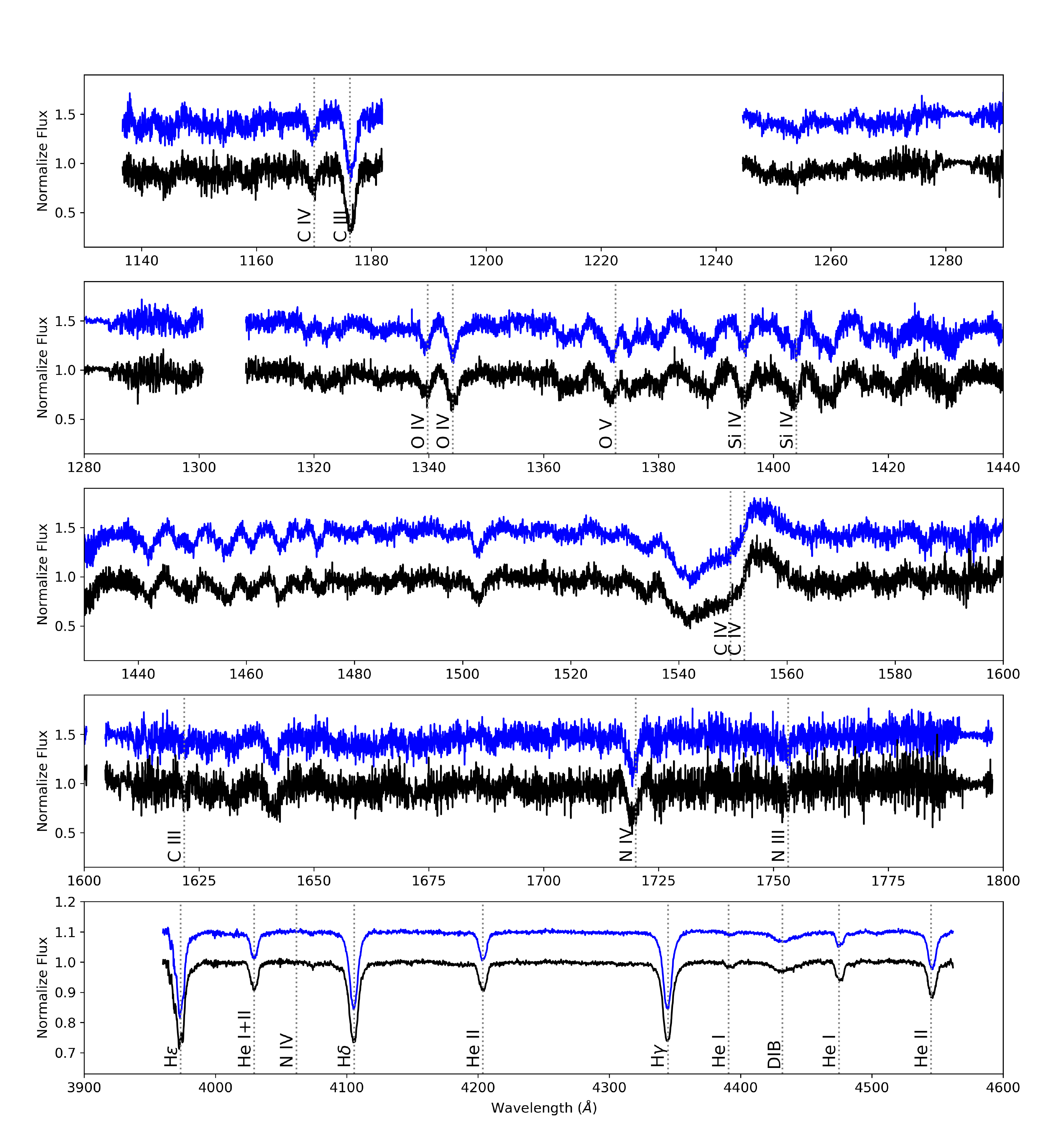}
      \caption{UV and optical disentangled primary (top) and secondary (bottom) component spectra for VFTS 352 with prominent diagnostic lines labeled.  The primary spectra have been vertically shifted for readability. 
              }
         \label{Disentangle_comparison_opuv}
   \end{figure*}
   \subsection{Removal of Nebular Lines}
   
      The second step of the disentangling procedure involves the removal of the nebular lines.  Using the radial velocity curve and the results of the three component disentangling from the previous step, we reconstruct each of the 32 original spectra using the spectra of the primary component, secondary component, and the spline fit of the third component.  Then, for each phase, we subtract the reconstructed spectra from the original spectra.  Since the splines do not contain the nebular lines, the only signals seen in the residuals of the subtraction should be from the nebular lines.  Since the nebular contributions are not constant in strength across the 32 original spectra, the final nebular line profiles generated in the previous section need to be scaled for each spectra.  Since the nebular lines are so sharp, each spans at most 3 or 4 pixels, slight wavelength shifts can affect the line profile.  Thus, we perform a least squares minimization to fit the nebular line profile to the residuals of the subtraction by leaving the scale factor and wavelength shift as free parameters.  Using the results of the minimization, we subtract the fitted nebular line profiles from the original spectra.  The results are the final spectra that we use for the two component disentangling.
      
   \subsection{Two Component Disentangling}

      The third and final step in the disentangling procedure is the two component disentangling.  The final spectra resulting from the nebular line removal are used as the input spectra for the two component disentangling.  The orbital parameters and Doppler velocities are set to those derived from \citet{Almeida2015} and are held constant, and the light ratios are calculated from the photometric data.  To avoid numerical issues, the spectra are disentangled in $\sim$20 \AA-wide sections for the UV data and $\sim$100 \AA-wide sections for the optical and the resulting spectra are stitched back together after disentangling. Due to the width of the Lyman alpha line, we remove it from our analysis, as it could not be disentangled in sections as per our standard methodology. The final results of the disentangling procedure consist of two spectra, one for each component, which are displayed on Fig.~\ref{Disentangle_comparison_opuv}.   After disentangling, the spectra have signal-to-noise ratios per resolution element of $\sim$600 and $\sim$530, respectively, in optical. The UV signal-to-noise ratios are $\sim$23 and $\sim$21 for the primary and secondary stars respectively.  
      
      Reconstructing the phased spectra and comparing them to the observed spectra show minimal signal in the residuals above the level of the noise.  No definite structure can be seen in the residuals aside from small imperfections in nebular line removal indicating that the disentangling was successful. Slight variations in spectral line residuals from one epoch to another can be seen in \hea~\l4471 and may be related to the presence of the Struve-Sahade effect \citep{Struve1937, Sahade1959, Linder2007}. The amplitude of the effect is limited to about 1.5\%\ of the continuum and does not seem to be visible in other spectral lines.
      
      The disentangled spectra of the two components are very similar at first sight. The primary component however displays weaker \hea\ lines while the \heb\ lines for both components appear comparable.  Qualitatively, this implies that the primary component is hotter than the secondary. Additionally, the wings of the H$\delta$ lines are asymmetric in the secondary component while they appear symmetric in the primary. 

\begin{table}
\caption{Summary of the diagnostic line list. The main identifier is given in Col.~1. Since some lines contain blends, the components contributing to each lines are listed in Col.~2 while the adopted fitting range is provided in Col.~3.}
\centering 
\begin{tabular}{ccc}
\hline\hline
Line Identifier & Components & $\lambda$ Fitting Range\\
 & & (\AA)\\
\hline
H$\delta$ 			& H \textsc{i}, He \textsc{ii}, N \textsc{iii}, Si \textsc{iv}  	& 4088.3 -- 4117.5	\\
H$\gamma$ 			& H \textsc{i}, He \textsc{ii}, Si \textsc{iv}  					& 4325.4 -- 4360.0	\\
He \textsc{i} 4026 	& He \textsc{i}, He \textsc{ii}	 								& 4022.8 -- 4035.9	\\
He \textsc{i} 4387 	& He \textsc{i} 													& 4384.8 -- 4399.0	\\
He \textsc{i} 4471 	& He \textsc{i} 													& 4469.6 -- 4481.1	\\
He \textsc{ii} 4200 	& He \textsc{ii}, N \textsc{iii} 								& 4194.9 -- 4212.0	\\
He \textsc{ii} 4541 	& He \textsc{ii} 												& 4537.0 -- 4553.9	\\
C \textsc{iii} 1176 	& C \textsc{iii} 												& 1174.3 -- 1178.4	\\
C \textsc{iii} 1620 	& C \textsc{iii} 												& 1619.5 -- 1623.5	\\
C \textsc{iv} 1169  	& C \textsc{iv} 													& 1167.7 -- 1171.9	\\
C \textsc{iv} 1548  	& C \textsc{iv} 													& 1536.4 -- 1565.6	\\
C \textsc{iii} 4069 	& C \textsc{iii} 												& 4066.5 -- 4078.5	\\
N \textsc{iii} 1750 	& N \textsc{iii} 												& 1747.0 -- 1757.0	\\
N \textsc{iv} 1718  	& N \textsc{iv} 													& 1715.4 -- 1723.0	\\
N \textsc{iv} 4058  	& N \textsc{iv} 													& 4055.5 -- 4067.5	\\
O \textsc{iv} 1340  	& O \textsc{iv} 													& 1337.4 -- 1347.0	\\
O \textsc{v} 1371   	& O \textsc{v} 													& 1369.0 -- 1373.5	\\
Si \textsc{iv} 1393 	& Si \textsc{iv} 												& 1393.0 -- 1397.0	\\
Si \textsc{iv} 1402 	& Si \textsc{iv} 												& 1402.0 -- 1405.9	\\
 \hline
\end{tabular}
\label{table:lines}
\end{table}

\section{Atmospheric Analysis} \label{sect: atm}
	
	In order to determine the atmospheric and wind parameters, as well as the surface abundances for each component star, we adjust the disentangled spectra with synthetic spectral line profiles generated by the stellar atmosphere code \textsc{fastwind}. \textsc{fastwind} is a 1D NLTE radiative transfer code for single stars that accounts for the expanding atmosphere and is suitable for the analysis of OBA stars \citep{Puls2005}.  Given a set of physical and atmospheric parameters, \textsc{fastwind} computes the model atmosphere and wind structure as well as spectral line profiles from selected (so-called "explicit") elements, in our case H, He, C, N, O, Si and P \citep{Puls2005, Rivero-Gonzalez2012}.  \textsc{fastwind}, however, does not have any fitting capabilities so a genetic algorithm is wrapped around it \citep{Charbonneau1995}. This technique, originally developed by \citet{Mokiem2005}, allows for an efficient exploration of the 11-dimension parameter space that is relevant for our specific objects.

	\subsection{Genetic Algorithm}
		As the name suggests, genetic algorithms mimics how natural selection works in nature.  Given a population, only the fittest individuals survive and reproduce.  The offspring  then have a combination of the genes from the parents and again only the fittest members of the new generation will survive and reproduce.  This cycle continues, eventually leading to a population of individuals that are significantly fitter than the original population.  This concept can be applied to create an algorithm that works in the same way.  In a genetic algorithm, the parameters used to calculate the model are analogous to the genes and the chi square of the fit is used to calculate a fitness metric.  The fitness metric $F$ is defined as
	\begin{equation}
      F \equiv \left( \sum_{i}^{N} \chi_{\mathrm{red}, i}^2 \right) ^ {-1} 
   	\end{equation}
where $N$ represents the total number of spectral lines being fit over and $\chi_{\mathrm{red}, i}^2$ represents the reduced chi square of the $i^\mathrm{th}$ spectral line \citep{Mokiem2005}.  The reduced chi square is discussed in further detail in Section \ref{errorcalc}.  

An initial population is created randomly across the parameter space.  The parameters from the models with the highest fitness are combined to form the next generation.  In addition, transcription errors and mutations are introduced (i.e. switching two digits, deleting a digit, etc.) allowing the algorithm to efficiently explore the parameter space and find a best fit solution \citep{Charbonneau1995, Mokiem2005}.  This method has already been used to analyze the optical spectra of other O-type stars from various instruments  \citep{Mokiem2006, Mokiem2007, Tramper2011, Tramper2014, Oscar2017}, however this is the first time that it is applied to UV and optical data simultaneously.
		
		We make several modifications and improvements to the genetic algorithm used in previous studies, however the fitting method and framework remain the same.  
		
		\begin{itemize}
	\item \textbf{\textsc{fastwind}:}
		The first major update is we use the most recent available version of \textsc{fastwind} \citep[version 10.3;][]{Sundqvist2018}.  With this update, we utilize a larger number of explicit elements than previously, which includes hydrogen, helium, carbon, nitrogen, oxygen, phosphorus and silicon (for nitrogen, we use the model atom from \citet{Rivero-Gonzalez2012}, and for C, O, Si, and P, we use the database from \citet{Pauldrach2001}).   Additionally, \textsc{fastwind} now supports treatment of UV lines. 
		
	\item \textbf{Line list:}  we modified the line list to include these new elements as well as the extended wavelength range.  Some lines used in previous studies were removed as they were either not present or outside of our wavelength coverage.  Our full line list can be found in Table \ref{table:lines} and include 24 spectral lines of H, He, C, N, O and Si spread over 19 spectral regions.
	
		\item \textbf{Microturbulence:} 
	We adopt a new depth-dependent microturbulence prescription in the formal solution, which now uses a value that increases linearly with wind velocity from a fixed photospheric value:
	\begin{equation}
		v_\mathrm{turb} = \mathrm{max}(v^\mathrm{ph}_\mathrm{turb}, 0.1v)
      \label{microturb_eq}
	\end{equation}
where $v^\mathrm{ph}_\mathrm{turb}$ is the photospheric microturbulent velocity and $v$ is the wind velocity. This prescription has been chosen to simulate the effects of the inhomogeneous wind-structure on the P-Cygni line profiles, in particular the observed black troughs and the blueward extension of such profiles beyond the terminal wind speed \citep[see, e.g.,][]{Hamann1980, Lucy1982, Prinja1990, Puls1993}

		\item \textbf{Adjusted parameters:}  we  expand our parameter set, which now contains 11 fitting parameters.  With the addition of UV lines, we are able to directly constrain wind parameters such as the terminal wind velocity and the exponent of the wind acceleration law.  Furthermore, with the addition of more explicit elements we can now attempt to constrain the abundances of carbon, nitrogen, oxygen and silicon.  Our parameters and the corresponding fit ranges can be found in Table \ref{table:params} and more details on each individual parameter and how they qualitatively affect the final synthetic spectra can be found in Appendix \ref{sect: atm-param-app}.  Appendix \ref{sect: atm-param-app} also discusses other parameters that affect the final synthetic spectral profile, for which we do not fit.  These parameters include microturbulence, macroturbulence, the inclusion of X-rays and clumping.
		
		\item \textbf{Luminosity anchor:} 
	 Previous uses of this genetic algorithm approach used either the $V$- or $Ks$-band magnitudes as  luminosity anchors to compute the stellar radii. Here we adopt the radii of the components as derived from the light curve analysis of  \citet{Almeida2015}.
		
		\item \textbf{Population size:} Most previous works used a population size of 78 individuals.  Because of the  larger set of parameters explored here, we increased our population size to 238 individuals. This values was chosen after several tests investigating the convergence of the fitting process. With such a population size, most fits converged within 50 to 70 generations while we let the algorithm run for about 170 generations to ensure sufficient  mapping of the global optimum. This is indeed needed  to derive proper confidence intervals on the fitted parameters as described in Sect.\ \ref{errorcalc}. 

	\end{itemize}

\begin{table}
\caption{Summary of the GA parameter set used for fitting.}
\centering 
\begin{tabular}{cccc}
\hline\hline
\multicolumn{2}{c}{Parameter} & Symbol & Fitting Range\\
\hline
Effective temperature				& $\left(\mathrm{K}\right)$							& $T_\mathrm{eff}$  			& 39000 -- 47000 \\
Surface gravity		 				& $\left[\mathrm{cm\ s}^{-2}\right]$					& log $g$  					& 3.5 -- 4.5	 \\
Projected rotational velocity 		& $\left(\mathrm{km\ s}^{-1} \right)$				& $v$ sin $i$ 				& 200 -- 400	 \\
Mass loss rate 						& $\left[\mathrm{M}_\odot\,\mathrm{yr}^{-1}\right]$	& $\log \dot{M}$ 			& $-$8.0 -- $-$6.0 \\
Velocity field exponent				& 													& $\beta$ 					& 0.5 -- 4.0	 \\
Terminal wind speed 					& $\left(\mathrm{km\ s}^{-1} \right)$				& $v_\infty$ 				& 500 -- 4000 \\
He surface abundance 				&													& $\varepsilon_\mathrm{He}$ 	& 0.05 -- 0.3 \\
C surface abundance 					&													& $\varepsilon_\mathrm{C}$ 	& 6.0 -- 9.0 	\\
N surface abundance 					&													& $\varepsilon_\mathrm{N}$ 	& 6.0 -- 9.0 	\\
O surface abundance  				&													& $\varepsilon_\mathrm{O}$	& 6.0 -- 9.0 	\\
Si surface abundance  				&													& $\varepsilon_\mathrm{Si}$ 	& 6.0 -- 9.0 	\\
 \hline
\end{tabular}
\label{table:params}
\end{table}

	\subsection{Error Calculation} \label{errorcalc}
		We use the same error calculation techniques described in \citet{Tramper2011, Tramper2014} and \citet{Oscar2017}.  The first step is to normalizing the overall $\chi^2$ values such that the lowest $\chi^2$ satisfies $\chi_\mathrm{red}^2 = 1$.  This implicitly assumes that the best model provides a satisfactory fit to the data.  This is done to ensure that the final error bars are not influenced by over- or under-estimation of the errors in the flux values. The next step is to calculate the probability ($P$) that the observed re-normalized $\chi_\mathrm{red}^2$ is not due to statistical fluctuation.  This probability is given by $P=1-\Gamma(\chi^2/2, \nu/2)$, where $\Gamma$ is the incomplete gamma function and $\nu$ is the degrees of freedom.  Models with $P \geq 0.05$ are considered acceptable models and the range of parameters of the accepted family of solutions are adopted as the 95\%-confidence interval.  As for all $\chi^2$-fitting methods, all models that satisfy $P \geq 0.05$ represent the data in a statistically indistinguishable way and are thus considered  \textit{statistically equivalent}. 

 \begin{figure}
   \centering
   \includegraphics[width=1\linewidth]{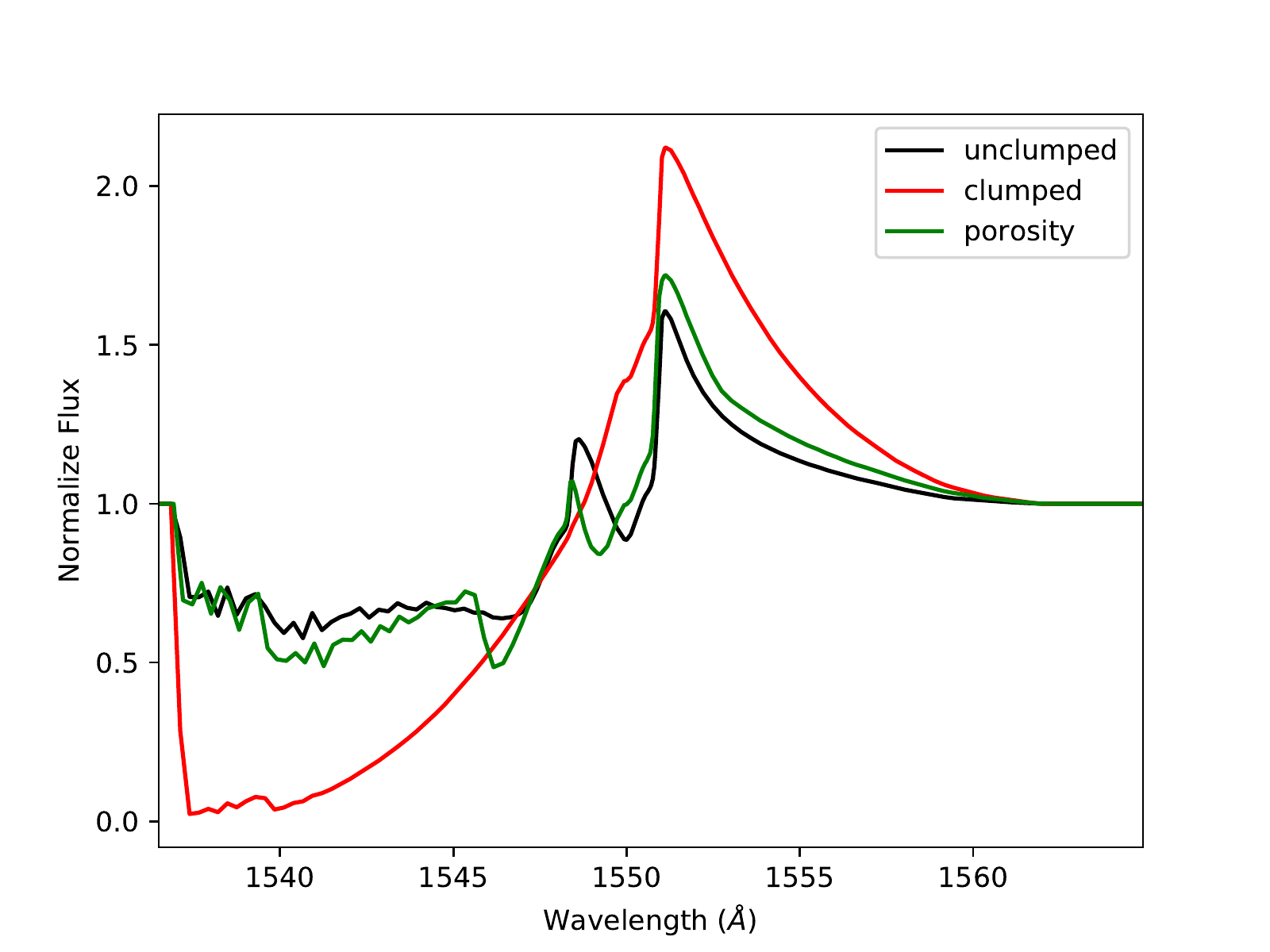}
      \caption{Two \textsc{fastwind} synthetic spectra with identical parameters except for the clumping factor, one with $f_\mathrm{cl}$ = 1 (black) and one with $f_\mathrm{cl}$ = 10 (red) are plotted to demonstrate the indirect effect of clumping (due to increased recombination) on the \cd~\l1548 resonance line.  Another synthetic spectrum (green) with the same parameters as the clumped model as well as porosity in velocity space included is plotted for comparison (using a velocity filling factor $f_\mathrm{vel}=0.3$) to show that the inclusion here of velocity-porosity counteracts the effects of optically thin clumping.  The parameters used for these computations can be found in the primary model with $\chi_\mathrm{min}^2$ column in Table \ref{table:ga_results}.
              }
         \label{c4_clump}
   \end{figure}   
   
	\begin{figure*}
   \centering
   \includegraphics[width=1\linewidth]{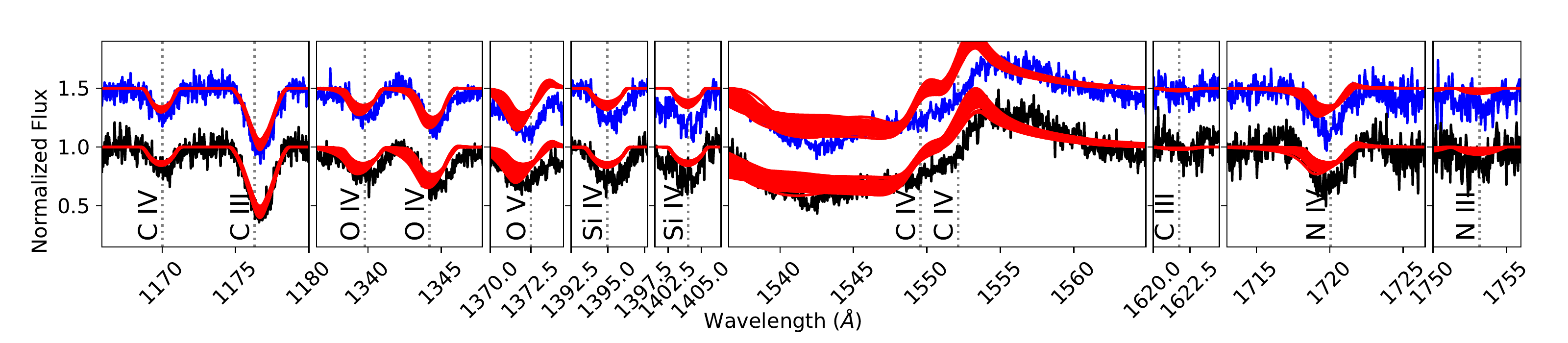}\\
   \includegraphics[width=1\linewidth]{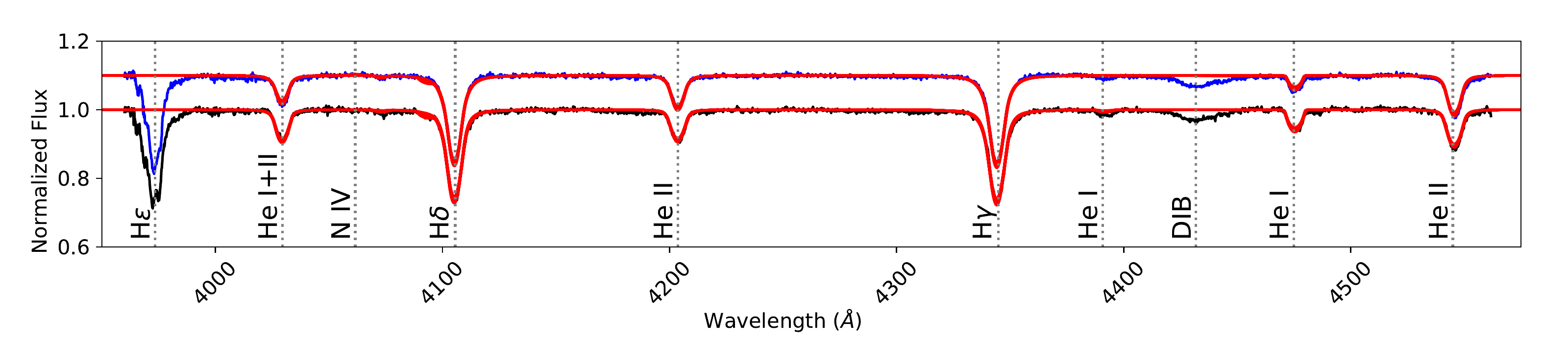}
      \caption{\textsc{fastwind} synthetic spectra (plotted in red) for each model in the family of statistically equivalent solutions are overplotted on the disentangled UV and optical data for the primary (blue) and secondary (black). 
              }
         \label{ga_fits}
   \end{figure*}
\subsection{Framework and Assumptions} \label{sect: errors}

In this section, we acknowledge two main assumptions that form the framework of the present work and we discuss the associated uncertainties and potential impacts on the derived parameters.   

\subsubsection{Non-spherical geometry}
One of the most prominent assumption probably concerns the geometry of the system.  As stated previously, VFTS 352 is an overcontact system so that the surface has roughly the shape of a peanut (see Figure 3 in \citet{Almeida2015}).  This non-spherical geometry is not properly accounted for in the disentangling and the atmosphere fitting.  Specifically, when disentangling, FDBinary assumes that the stars are well separated point sources and that the individual spectral signatures do not vary with phase, aside from the scaling due to light ratio variations.  The von Zeipel theorem suggests that the bridge of overcontact systems should be significantly cooler than the poles \citep{vonZeipel1924}, meaning that the spectral contribution from this region is probably very different from that of the poles.  Thus, we would expect to see slight temperature and surface gravity variations as a function of phase as possibly suggested by residual variation in, e.g.,  the \hea~\l4471 line as discussed above. From a PHOEBE model \citep{Prsa2016} corresponding to the light curve solution of \citet{Almeida2015}, we estimate temperature differences of the order of 5\%\ between the pole and the equator and 30\%\ between the pole and the bridge, however, the surface area subtended by these regions account for a very small portion of the total surface area and therefore these regions do not strongly affect the results of the disentangling.  These spectral differences are washed out by the disentangling procedure, so that the resulting disentangled spectra are effectively time- and space-averaged profiles across the surface of each star.  It should also be noted that the disentangling procedure does not distinguish between photospheric and wind lines, and applies the same radial velocity shifts and light ratios to both.

The non-spherical nature of the system also presents a problem when performing the atmosphere fitting.  \textsc{fastwind} is a one dimensional code, that implicitly assumes that the stars are spherical.  Furthermore, \textsc{fastwind} is designed for single stars, with the assumption that there is no incident intensity from outside sources at the outer boundary.  Neither of these conditions are met in this case, however it should be noted that there are no three-dimensional NLTE radiative transfer codes currently available.

\subsubsection{Clumping}
Another important source of uncertainty comes from the effect of clumping on the ionization balance. Given the parameters of this system, the increased density in the clumps is enough to cause a noticeable shift in the ionization balance of carbon, specifically, \ce\ recombines to form \cd. Figure \ref{c4_clump} illustrates the effect of optically thin clumping on the \cd~\l1548 resonance doublet. It  shows how the \cd\ lines become stronger, implying a lower carbon abundance for the same observed line strength. A similar effect on the \cd\ line would also result from increasing the mass-loss rate, but in this case other lines in the UV (especially the \nd~\l1720) become too strong. On the other hand, the \cd\ doublet is expected to weaken in strength if light-leakage effects associated with porosity in velocity space are considered, as compared to the optically thin clumping model. Figure \ref{c4_clump} illustrates this, showing the \cd\ doublet for a comparison model including velocity-porosity using a velocity filling factor $f_\mathrm{vel}=0.3$ \citep[see ][for details]{Sundqvist2018}. Considering the quite large effect found on the \cd\ lines, these clumping-effects certainly need to be investigated in more detail in future work.

\begin{table*}
\caption{Results of the 11-parameter genetic algorithm fit of \textsc{fastwind} models to  the primary and secondary components assuming an unclumped wind.  The model corresponding to the lowest $\chi^2$ as well as the 95\%-confidence intervals are given.}
\centering 
\begin{tabular}{llcccccc}
\hline\hline
& & & \multicolumn{2}{c}{Primary} & & \multicolumn{2}{c}{Secondary}  \\
\multicolumn{2}{c}{Parameter} & & Model with $\chi_\mathrm{min}^2$ & Confidence interval & & Model with $\chi_\mathrm{min}^2$ &  Confidence interval \\
\vspace*{-3mm}\\
\hline
\vspace*{-3mm}\\
$T_\mathrm{eff}$				& $\left(\mathrm{K}\right)$							& & 44200		& 42850 -- 45550			& & 40750 	& 40600 -- 41550 		\\
log $g$						& $\left[\mathrm{cm\ s}^{-2}\right]$					& & 4.14 		& 4.09 -- 4.24			& & 3.90 	& 3.80 -- 4.00			\\
$v$ sin $i$					& $\left(\mathrm{km\ s}^{-1} \right)$				& & 268.0 		& 240.0 -- 284.0			& & 296.0 	& 278.0 -- 310.0			\\
$\log \dot{M}$				& $\left[\mathrm{M}_\odot\,\mathrm{yr}^{-1}\right]$	& & $-$7.1 		& $-$7.25 -- $-$6.95		& & $-$7.05 	& $-$7.50 -- $-$7.00		\\
$\beta$						&													& & 3.30 		& 2.30 -- 4.00			& & 1.55 	& 1.45 -- 2.95			\\
$v_\infty$					& $\left(\mathrm{km\ s}^{-1}\right)$					& & 2300 		& 2000 -- 2900			& & 2600 	& 2200 -- 4000			\\
$\varepsilon_\mathrm{He}$	& 				 									& & 0.10 		& 0.085 -- 0.13			& & 0.08 	& 0.07 -- 0.10			\\
$\varepsilon_\mathrm{C}$		& 													& & 7.70 		& 7.40 -- 8.00			& & 7.25 	& 7.05 -- 7.55			\\
$\varepsilon_\mathrm{N}$		& 													& & 6.40 		& 6.10 -- 7.55			& & 6.20 	& 6.00 -- 7.35			\\
$\varepsilon_\mathrm{O}$		& 													& & 8.35 		& 7.45 -- 8.65			& & 8.00		& 7.45 -- 9.00			\\
$\varepsilon_\mathrm{Si}$	& 													& & 6.95 		& 6.00 -- 7.45			& & 6.50 	& 6.00 -- 7.15			\\
 \hline
\end{tabular}
\label{table:ga_results}
\end{table*}


\section{Atmospheric Analysis Results} \label{sect: results}
	After disentangling, we separately ran the genetic fitting algorithm to adjust \textsc{fastwind} profiles to the primary and secondary components and determine their stellar and wind parameters. The results are given in Table \ref{table:ga_results} and displayed in Fig. \ref{ga_fits}. Figures \ref{ga_fitness_plots-vfts352a} and \ref{ga_fitness_plots-vfts352b} in the Appendix, show the $\chi^2$ distribution projected on the individual parameter axis. Here we first briefly discuss the results. The evolutionary status of VFTS~352 will be discussed in Sect.~\ref{sect: evol}. 
	
		
	
		The effective temperatures for both the primary and the secondary are in the ranges of 43000 -- 45500~K and 40500 -- 41500~K, respectively.  The derived values for the projected rotational velocities (in the range of 240 to 310~\kms, depending on the object) are  in  good agreement with expectation from tidal locking ($v_\mathrm{eq} \sin i \sim 270$~\kms). The derived surface gravity of $\log g \approx 3.8$ to 4.2 for both components are normal for main sequence O-type dwarfs.  The mass loss rates of  $\log \dot{M} \approx -7.0$ (in units of \msun~yr$^{-1}$) is however lower than predicted from \citet{Vink2001} by a factor of about 6 for the primary and 2 for the secondary. This can be related to the fact that, with bolometric luminosities of about $\log L_\mathrm{bol} / L_\odot \approx 5.2$, the VFTS~352 components are at the limit of delivering enough UV flux to drive a strong stellar wind \citep{Martins2005, Muijres2011}.   The low mass loss rate is reinforced by the fact that the \cd~\l1548 wind line is not saturated, as it would be with a higher mass loss rate.  The $\beta$ is higher than expected with values in the ranges of 2.5 to 4 for the primary and 1.5 to 3 for the secondary.  The helium abundances for both components are approximately 0.1, while the nitrogen and silicon abundances are smaller than 7.5, the carbon abundance between 7 and 8 and the oxygen abundance between 7.5 and 9.
	
		The photometric analysis of VFTS 352 by \citet{Almeida2015} concluded that the temperature of the primary and secondary are too hot for their dynamical masses when considering mixing from rotation alone (42540 K $\pm$ 280 and 41120 K $\pm$ 290 for the primary and secondary respectively).  Our results  also showed increased temperatures, qualitatively agreeing with the conclusions of \citet{Almeida2015}.  The primary temperature that we derived  is slightly higher than what was found in \citet{Almeida2015}.  The temperature of the secondary is however consistent with their photometric analysis. The only other parameter in our data set that was constrained by \citet{Almeida2015} is the surface gravity, with values of  4.18 $\mathrm{cm\ s}^{-2}$ $\pm$ 0.01 both the primary and secondary.  The surface gravity that we derived for the primary (log $g$ = 4.09 -- 4.24) is consistent with the photometric analysis, however the one derived for the secondary (log $g$ = 3.80 -- 4.00) is smaller by about 0.2 dex. It is known that {\sc fastwind} tends to provide lower surface gravities than \textsc{cmfgen} \citep{Hillier1998}, by typically 0.1 dex \citep{Massey2013}. Accounting for this difference would resolve the apparent discrepancy.

	\subsection{Robustness of the results}
	
		Several tests were conducted to assess the robustness of the resulting parameters.  Since our parameter set did not include clumping or X-rays, we conduct two separate fitting runs for each component where we include these parameters. Additionally, we compute several other GA runs where we hold certain parameters constant or remove certain lines to determine whether and how these affect the final solution.  The obtained results are detailed below. Each fitting run uses the same base setup, parameters and line list as detailed in Section~\ref{sect: atm} with the exception of the changes explicitly stated in the following.
		
	\begin{itemize}

		\item	\textbf{Inclusion of X-rays} - As discussed in Appendix~\ref{sect: atm-param-app}, X-rays affect the ionization balance.  We conducted a fitting run with the X-rays option turned on in  \textsc{fastwind}. There are five X-Ray parameters needed for the \textsc{fastwind} calculations and they are described in \citet{Carneiro2016}.  For this run, we adopt the following parameter values: $f_\mathrm{X} = 0.06,\ u_\infty = 400,\ \gamma_\mathrm{x}=0.75,\ m_\mathrm{x} = 20,\ R^\mathrm{input}_\mathrm{min}=1.5 $, which corresponds to an X-ray luminosity of $L_\mathrm{x}/L_\mathrm{bol} = 2 \times 10^{-8}$.  An X-ray luminosity of $L_\mathrm{x}/L_\mathrm{bol} = 1 \times 10^{-7}$ is a typical value for a main sequence star in this mass range, however one might expect a lower X-ray luminosity for a system with weaker winds for a given luminosity (as is the case in this system) since the X-ray luminosity scales directly with the star's mass loss rate \citep{Owocki1999}.  The addition of X-rays caused the temperature for both the primary and secondary to drop by around 1000~K but they still agree within error.  The surface gravity and the mass loss also decrease slightly for both but remain within error.  Additionally, the overall confidence intervals for $\beta$ and for the oxygen abundance increase by around 0.5, both for the primary and secondary however they still largely overlap with the confidence intervals given in Table~\ref{table:ga_results}. The nitrogen and carbon abundance are left unaffected. The resulting best-fit line profiles are left unchanged, but with very small variations in the \cd~\l1548 line. We thus concluded that X-ray production in the stellar winds do not critically affect our results, except maybe of a $\sim$1000~K shift in $T_\mathrm{eff}$. \\
		
		\item\textbf{Inclusion of clumping} - As discussed earlier, clumping can affect both the shapes and strengths of the wind lines and can further alter the ionization balance in the wind.  We conduct a fitting run with a fixed clumping factor $f_\mathrm{cl}$ = 10 and assuming optically thin clumps. The addition of such clumping lowers the temperature by about 500~K but remains within error in the primary star. It raises the temperature of the secondary by about 500~K, though again this value is well within error.  Moreover, there is an increase in the $\beta$ parameter to well above 3.5 for both the primary and the secondary.  Additionally, the carbon abundance drops to around 6.8 for both the primary and the secondary,  i.e., significantly smaller than the confidence interval listed in Table~\ref{table:ga_results}. The reduction in carbon abundance is driven by the need to have an unsaturated \cd~\l1548 line. All other parameters, including the mass-loss rate, are left unchanged. \\
		
		\item\textbf{Fixed $\beta$} - The $\beta$ values that we determined for both the primary and secondary determined are  higher than predicted. For O-type main sequence stars, $\beta$ values between 0.8 and 1.0 are usually expected \citep[e.g.,][]{Puls1996, Muijres2011}.  We conducted fitting runs with $\beta = 0.9$ to see if the high beta found here has any significant effects on the final parameters.  For both the primary and the secondary, the range of acceptable temperatures narrows and drops by around 500 - 1000~K but remains within error.  Additionally, the surface gravity drops by 0.1 for the primary -- still within error -- and by 0.2 for the secondary, which is no longer within  error.  The mass loss rate decreases by approximately 0.2~dex for the primary but remains the same for the secondary. The oxygen abundance increases for both the primary and the secondary by about 0.5 but these values are still consistent within error. The range for the carbon abundance narrows slightly but not significantly.  One significant difference however is that the projected rotation rates increase by about 50~\kms\ and are no longer in agreement with values listed in Table~\ref{table:ga_results}, nor with the  projected rate of 270~\kms\ that is expected from tidal locking. A visual inspection indeed reveal that the optical lines suffer too heavily from rotational broadening, so that this lead to a significantly worse fit. As a final check,  we  also conducted two additional GA runs using $\beta=0.9$, however now also (i) removing the \od~\l1340, as we have found it to be particularly sensitive to $\beta$, and (ii) removing both \od~\l1340  and \cd~\l1548, to see whether this would affect the derived abundances for carbon and oxygen. However, these two runs turn out to be almost identical to the ones where these lines are included.\\
		
		\item\textbf{Fixed stellar wind properties} - Our analysis finds mass loss rates that are almost an order of magnitude lower than predicted by \citet{Vink2001}. We thus run the fit fixing all of the wind parameters such that they follow the relations described in \citet{Vink2001}, specifically, we set the $\beta$ to 0.9, the mass loss rate to $-$6.46 and $-$6.77 for the primary and secondary respectively and the terminal wind velocity to 2986 and 2991~\kms, respectively.  It leads to a significantly worse fit of the \cd~\l1548, lower $T_\mathrm{eff}$ (by 1500~K), higher gravity (by 0.1~dex) as well as lower abundances of N, C and O (by $\sim$1.0 dex) for the primary star. The situation is less dramatic for the secondary, with similar fit quality for all lines, a higher  $T_\mathrm{eff}$ (by 900~K) and similarly lower N, C and O abundances. Fixing the stellar wind properties to the prescription of \citet{Vink2001}  actually bring the largest differences in our entire analysis.  This strengthens the hypothesis that the components of VFTS 352 have weak winds \citep[e.g.][]{Martins2005, Muijres2011}, for which the \citet{Vink2001} prescription is known to break down.  Implications for the evolutionary state of the object are discussed in the next section. This test demonstrates the importance of adjusting the abundances and wind parameters simultaneously. Without including the wind parameters in the fit, large systematic errors in the abundances are introduced.\\
		
		\item\textbf{Non-depth-dependent Microturbulence} - We also compare our prescription for the depth-dependent microturbulent velocity with the radially fixed microturbulence prescription used in previous studies \citep{Mokiem2007, Tramper2011, Tramper2014, Oscar2017}. In these studies, the microturbulence is not depth-dependent and two separate microturbulent velocities are used, one for the computation of the NLTE occupation numbers and the atmosphere and wind structure, and a second one for the formal solution used to compute the synthetic spectra.  As a test, we set the microturbulence in the NLTE calculations to 15 km $\mathrm{s}^{-1}$ and we leave the one used in the formal solution as a free parameter.  The resulting $v_\mathrm{turb}$ is $50^{+5}_{-10}$~\kms.  While most parameters remain the same, some do change when the non-depth-dependent $v_\mathrm{turb}$ prescription is used.  For example, the temperature of the primary is lower by 1500 and 500~K for the primary and secondary. Moreover, $\beta$ for both the primary and secondary now become even higher, with values between 3 and 4.  Additionally, while there is some overlap in the confidence intervals, the carbon abundance for the primary is lowered by approximately 0.4 dex, suggesting some degeneracy between microturbulence and abundances. Qualitatively, the line profiles from the non-depth-dependent microturbulence actually gave a closer match to the observed spectra as compared to the line profiles from the depth-dependent microturbulence. Specifically, for the latter several of the UV lines that are sensitive to the wind start turning into P-Cygni profiles, making the absorption lines appear slightly blue-shifted. \\

		\item \textbf{Other NLTE atmosphere codes} --
		We also compute \textsc{cmfgen} and \textsc{PoWR} \citep{Grafener2002, Hamann2003, Sander2015} models with the parameters adopted from the best fit solution listed in Table~\ref{table:ga_results} as a final test to see whether the parameters derived from \textsc{fastwind} are reasonable and consistent. Figure \ref{fcp_comparison} shows the comparison between the line profiles computed with these three codes. Overall, the three codes are in quite good agreement.  There are some slight deviations in some of the line profiles of the more complex wind lines such as C \textsc{iv} 1548. The \hea~\l4471 to \heb~\l4541 ratio computed by \textsc{cmfgen} and \textsc{fastwind} are in very good agreement while a fit with \textsc{PoWR} would have likely yielded a slightly lower temperature to increase \hea~\l4471. Both the \textsc{cmfgen} and the \textsc{PoWR} models provide an actual better fit to the line profiles affected by the adopted depth-dependent microturbulence prescription (see above), suggesting that the abundances that we derived are indeed optimum despite uncertainties linked to the latter prescription.
			\end{itemize}
			

   \begin{figure*}
   \centering
   \includegraphics[width=1\linewidth]{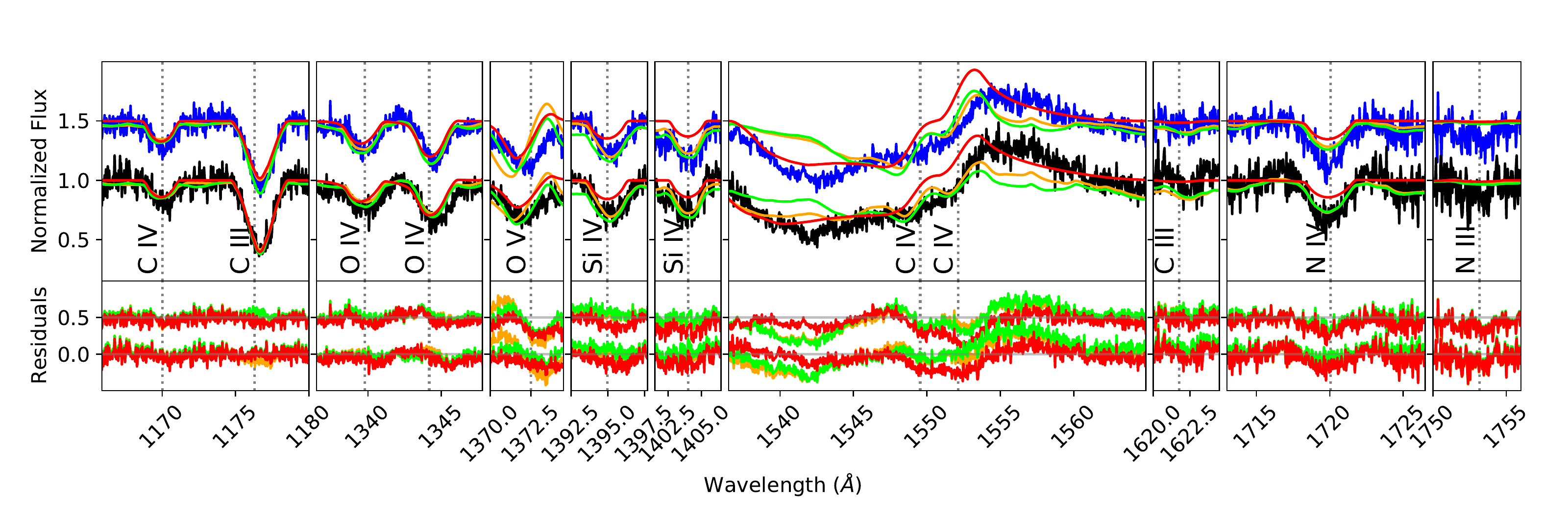}\\
   \includegraphics[width=1\linewidth]{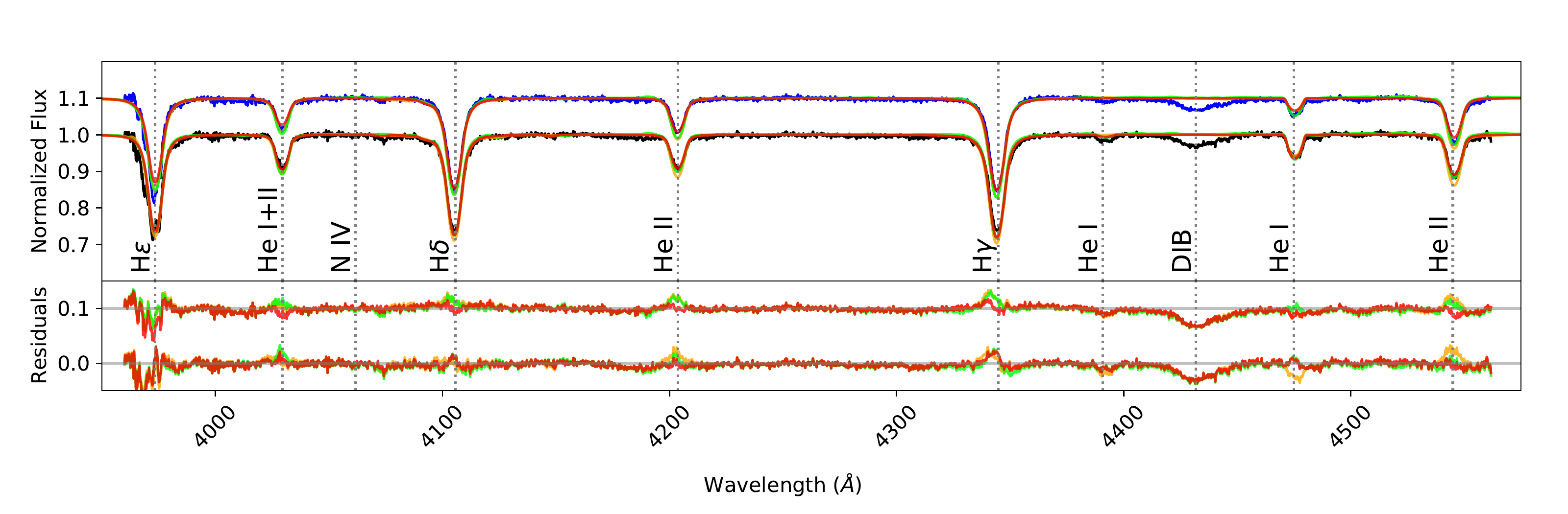}\\
      \caption{\textsc{fastwind}, \textsc{cmfgen} and \textsc{PoWR} synthetic spectra with the same input parameters (plotted in red, green and orange respectively) for the models with $\chi_\mathrm{min}^2$ are overplotted on the disentangled UV and optical data for the primary (top) and secondary (bottom) components. Residual plots showing observed minus model are included with the same color scheme.  For convenience zero-point residual lines are plotted in gray.
              }
         \label{fcp_comparison}
         
   \end{figure*}
		
		We also perform additional tests to ensure that the errors obtained are reasonable and that the values expected for certain parameters can be rejected with reasonable confidence.  To do this, we take the \textsc{fastwind} model with $\chi_\mathrm{min}^2$ for the primary and secondary and determine which rotation rate is needed to recover the current positions of the components on the HRD.  For single star evolution as described by \citet{Brott2011a}, very rapidly rotating initially 30 \msun\ stars recover the Hertzsprung-Russell diagram positions of the two components with a rotation rate of 512 \kms{}.  From these evolution models, we extracted the abundances of helium, nitrogen, carbon and oxygen that correspond with the HRD position and computed a series of additional {\sc fastwind} models where we change each of these abundance values to the predicted one to see how the spectrum changes. We do a similar exercise with $\beta$ and the mass-loss rate, where we set $\beta$ to 0.8 \citep{Muijres2011} and the mass-loss rate to that expected from the \citet{Vink2001} recipe.  Figures 9 and 10 show a comparison between the best fit family of solutions generated from the genetic fitting and the expected values detailed above for both the primary and secondary.
		
		These plots reveal that many of the theoretically predicted values result in spectral lines that are clearly incompatible with the observed profiles.  The expected mass loss rate for both the primary and secondary produce a \cd~\l1548 resonance line that is too strong in both the emission and absorption components.  In addition to the line shown, the expected mass loss rate also forces other wind lines in the UV to turn P-Cygni shaped. As pointed out above, a likely explanation for this is that the two stars lack the luminosity to develop a strong wind \citep[][e.g.]{Puls2008}. For galactic stars, \citet{Muijres2011} predicted that wind driving becomes strongly decreased at $\log L_\mathrm{bol} / L_\odot \approx 5.2$ \citep[see also ][]{Lucy2012}.  As a result of their chosen physical treatment of wind driving, this is not accounted for (self-consistently) by \citet{Vink2001}. Winds in this low-luminosity regime are referred to as ‘weak winds’.  For LMC stars the boundary between strong and weak winds will be at a somewhat higher luminosity, i.e. above the luminosities of VFTS 352 (of $\log L_\mathrm{bol} / L_\odot \approx 5.2$). Consequently it is expected that the components of VFTS 352 have weak winds. 
		
		The expected $\beta$ drastically changes the shape of the \cd~\l1548 line, causing it to disagree strongly with observations.  For the helium abundance predicted by \citet{Brott2011a} the lines of this species are too deep, i.e. the actual He abundance must be lower.  The expected carbon abundance does not fit the \cc~\l1176 and \cd~\l1169 doublet.  The expected nitrogen abundance causes the \nd~\l1720 line to turn P-Cygni, which clearly does not fit with the observations.  The expected oxygen abundance is within the error range for the secondary but too low for the primary.  We conclude that these tests show that the components of VFTS\,352 do not comply with evolution predictions of single massive stars.
		
		\begin{figure*}
   \centering
   \includegraphics[width=1\linewidth]{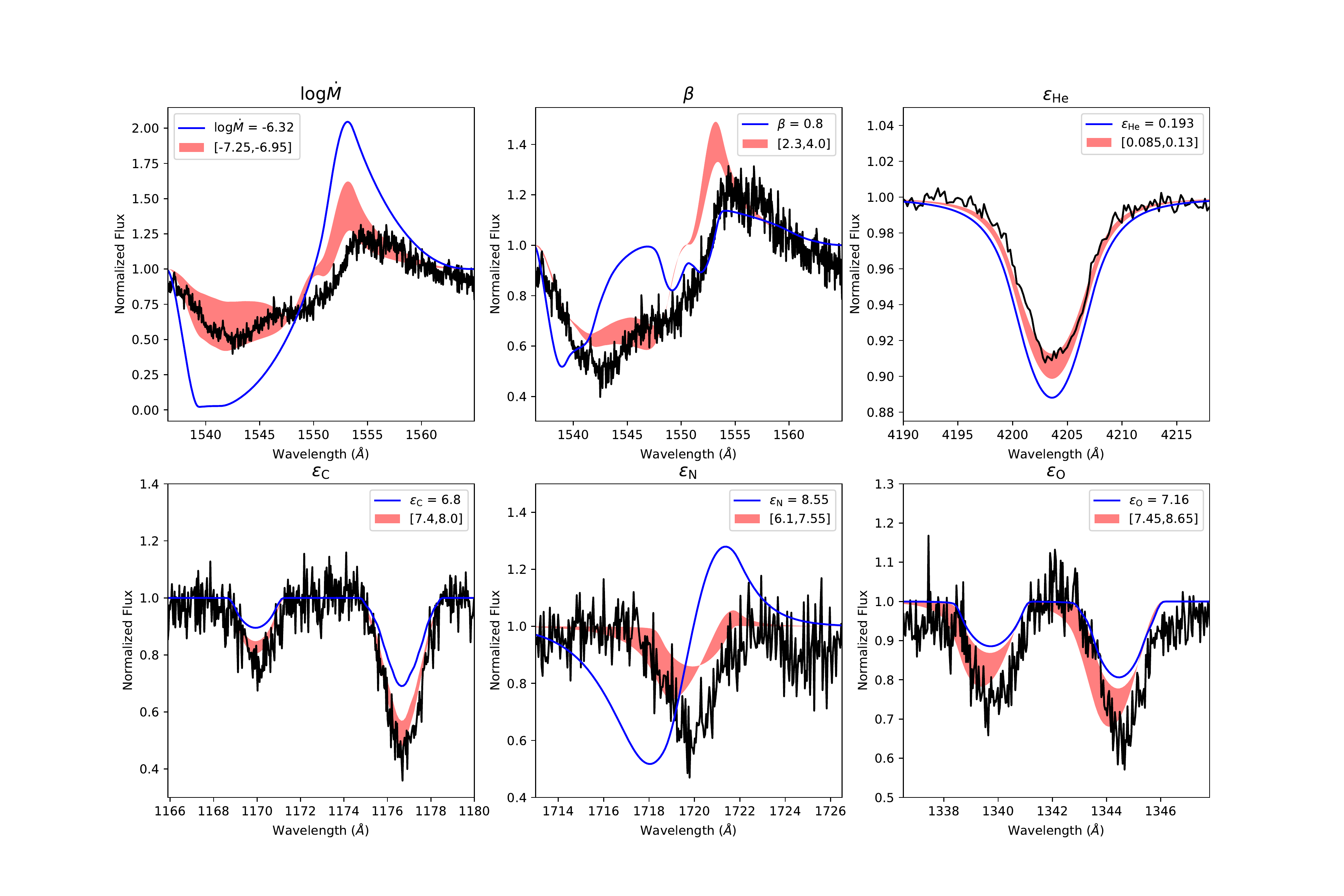}
  
      \caption{\textsc{fastwind} synthetic spectra representing the confidence intervals for 6 parameters (plotted as  red shaded areas) are overplotted on specific lines sensitive to each parameter in the disentangled UV and optical data for VFTS~352's primary component. An extra \textsc{fastwind} synthetic spectra is computed for each parameter using (i) the \citep{Vink2001} mass loss rate, (ii) $\beta$ of 0.8, and (iii-vi) the expected He, C, N, and O abundances corresponding to the derived effective temperature of the star (see text).
              }
         \label{error_validation-352a}
   \end{figure*}
		
		\begin{figure*}
   \centering
   \includegraphics[width=1\linewidth]{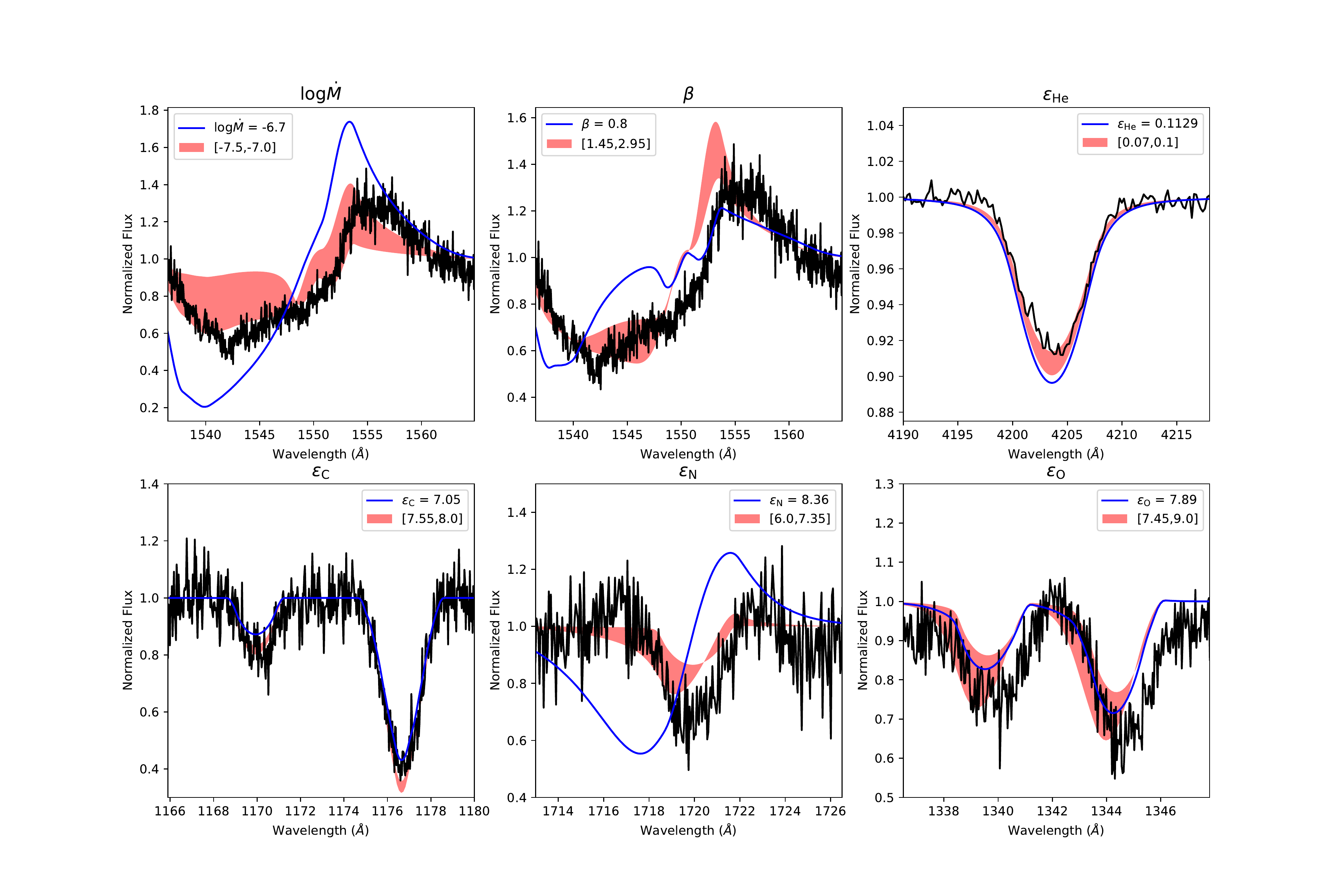}
      \caption{Same as Fig.~\ref{error_validation-352a} for the secondary component.
              }
         \label{error_validation-352b}
   \end{figure*}


	\begin{figure}
   \centering
   \includegraphics[width=0.9\linewidth]{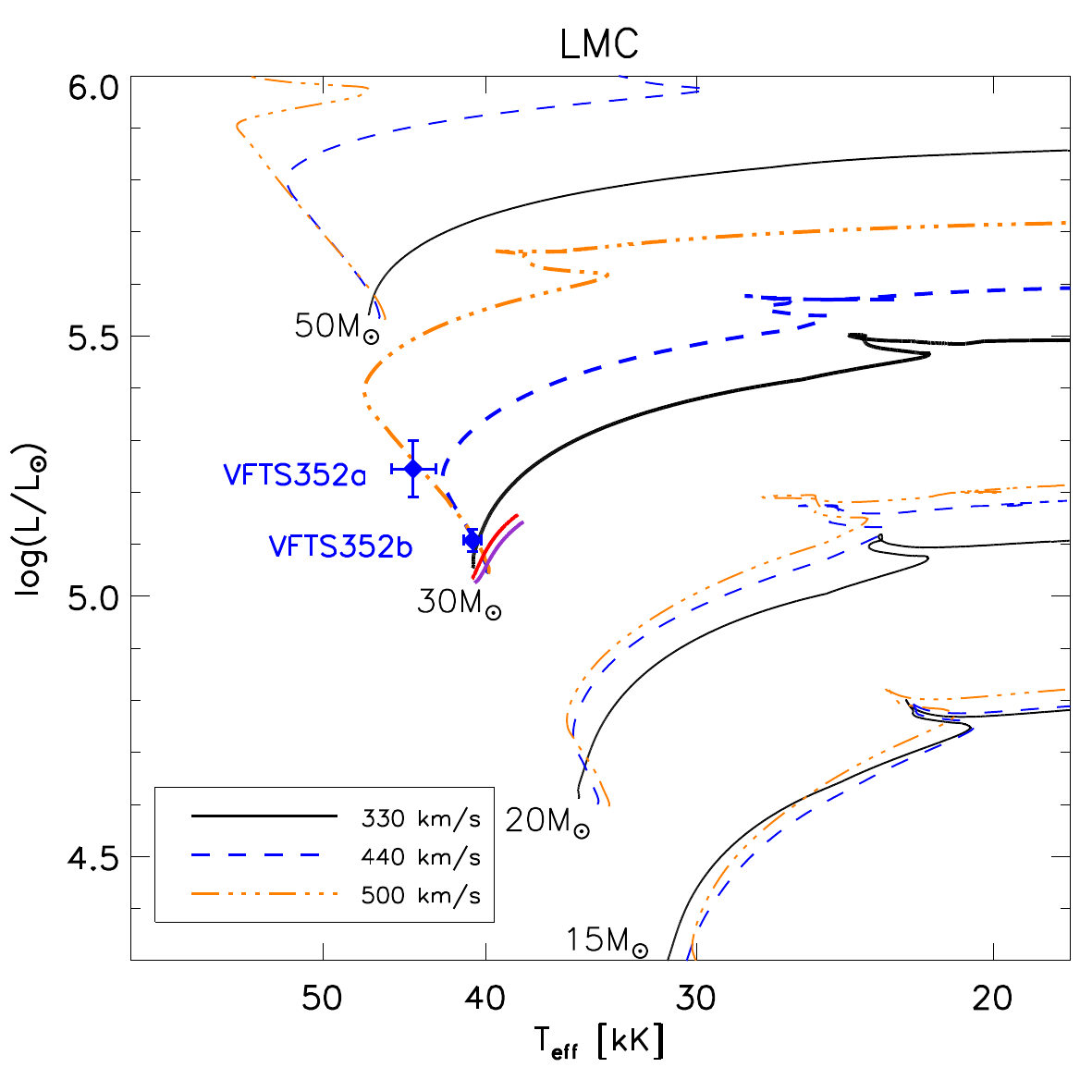}
      \caption{Location of VFTS 352 in the HR diagram. Single star LMC evolutionary tracks from  \citet{Brott2011a} for 15, 20, 30, and 50~\msun{}\ and three different initial rotation rates are indicated.  The best fit binary model and the enhanced-mixing binary models are plotted in purple and red respectively.
              }
         \label{hrd}
   \end{figure}
   
	\section{Evolutionary Status} \label{sect: evol}

		
	Currently no appropriate grid of evolutionary models has been published to which we can compare our observations, however computing such a grid is beyond the scope of the present paper. Here we compare our observational constraints to single star evolutionary tracks as well as a set of MESA binary evolution models for qualitative comparison.
	
	According to mixing theory, higher rotation rates correspond to more mixing.  Therefore, the rotation rate can be used as a proxy for measuring the expected efficiency of the mixing processes in a star.  By comparing our results with evolutionary models at different rotation rates, we can discuss whether more or less mixing (higher or lower rotation rate) is required to reproduce the abundances and temperatures observed. 
	
	\subsection{Location in the HRD} \label{sect: hrd}
	
	We first compare the effective temperatures that we derived to models from \citet{Brott2011a} which account for rotationally induced mixing through shear and Eddington Sweet circulations.  
	Figure \ref{hrd} shows the location of VFTS~352 primary and secondary components in the Hertzsprung-Russell diagram. For a 30~$\mathrm{M}_\odot$ star at LMC metallicity with initial abundances of 7.75, 6.90, 8.35 and 7.20 for carbon, nitrogen, oxygen and silicon respectively \citep{Kurt1998, Hunter2007}, rotating at 332 km $\textrm{s}^{-1}$, the expected zero age main sequence (ZAMS) temperature is $\sim40\,000$~K \citep{Brott2011a}. As already discussed by \citet{Almeida2015}, both stars lay to the left of their single-star evolutionary track given their masses and rotation rates. 
	The effect is most prominent for the primary star and is even stronger in the current analysis compared to \citet{Almeida2015}, with differences of at least 3000~K for the primary star. Significant additional mixing, e.g., more efficient rotational mixing than implemented in the models, is required for single star evolutionary models to reproduce the location of VFTS~352's two components in the HRD. Using enhanced rotation rates as a proxy for enhanced mixing, an initial rotation rate as high as  500~\kms{} would be required for VFTS~352a (Fig.~\ref{hrd}). This is an indication that the primary star could be evolving chemically homogeneously.

Since there is no suitable grid of binary models available (see however \citealt{Marchant2016}), we computed a set of binary-star models by varying initial periods (P=1.0-1.2\,days) and masses ($q=M_{2}/M_{1}=0.6-1$ and $M_{2}+M_{1}=56.4\,\mathrm{M}_{\odot}$, the currently-inferred total mass of the system), and assuming a rotational mixing factor of $f_\mathrm{c}$=0.033, which is a commonly adopted value \citep{Heger2000}. At the start of the calculation the stars begin with a uniform initial composition appropriate for young stars in the LMC \citep{Brott2011a}. For these calculations we used the stellar evolution package MESA (version 10398, \citealt{Paxton2011,Paxton2013,Paxton2015,Paxton2018}), with the physical assumptions from \citet{Marchant2016} for modeling the over-contact phase. Mass transfer is treated as conservative. The parameters used to compute our models are available as MESA inlist files \footnote{\url{https://github.com/orlox/mesa_input_data/tree/master/2016_double_bh}}.

We find the binary model from our grid that most closely matches the data is an equal-mass binary, with initial masses of $28.74+28.74\,\mathrm{M}_{\odot}$ and an orbital period of 1.07\,days. This system is initially in an over-contact configuration (i.e., both stars are larger than their Roche lobes), and it soon detaches and then once again reaches overcontact as the stars expand on the main sequence. There is no net mass transfer in this system since the stars begin with equal masses. The stars are in an over-contact configuration on the main sequence at $\approx1.8\,$Myr, at which stage this model has stellar masses, radii, rotational velocities, a mass ratio, and an orbital period that are roughly consistent with the observationally-inferred parameters. It does not, however, reproduce the observed luminosities or effective temperatures.  This calculation continues until the stars overflow their outer Lagrangian points, after which a merger can be expected to take place.  

We also tested an enhanced-mixing version of this model, keeping all other initial parameters the same and only changing the rotational mixing factor to $f_\mathrm{c} = 0.2$.  This model again predicts luminosities and temperatures lower than the observationally-inferred values for VFTS 352.  This may indicate more efficient mixing of helium in this system than in our models. The evolutionary tracks associated with the best fit binary model and the enhanced-mixing binary model can be found in Fig. \ref{hrd}.

	\subsection{Surface abundances} \label{sect: surf_abun}
	
	Given the high rotation rates and higher than expected effective temperatures, significant surface enrichment is expected. However, we detect no sign of strong He or N enrichment  while our data would allow us to do so (see e.g., Figs.~\ref{error_validation-352a} and \ref{error_validation-352b}). The derived abundances of C and O are possibly subject to uncertainties related to stellar wind physics. Due to the absence of strong optical diagnostics lines, our abundance diagnostics rely on the UV-portion of the spectrum. The derived Helium and Nitrogen abundances, however, should be more robust.

\begin{figure*}
   \centering
   \includegraphics[width=1\linewidth]{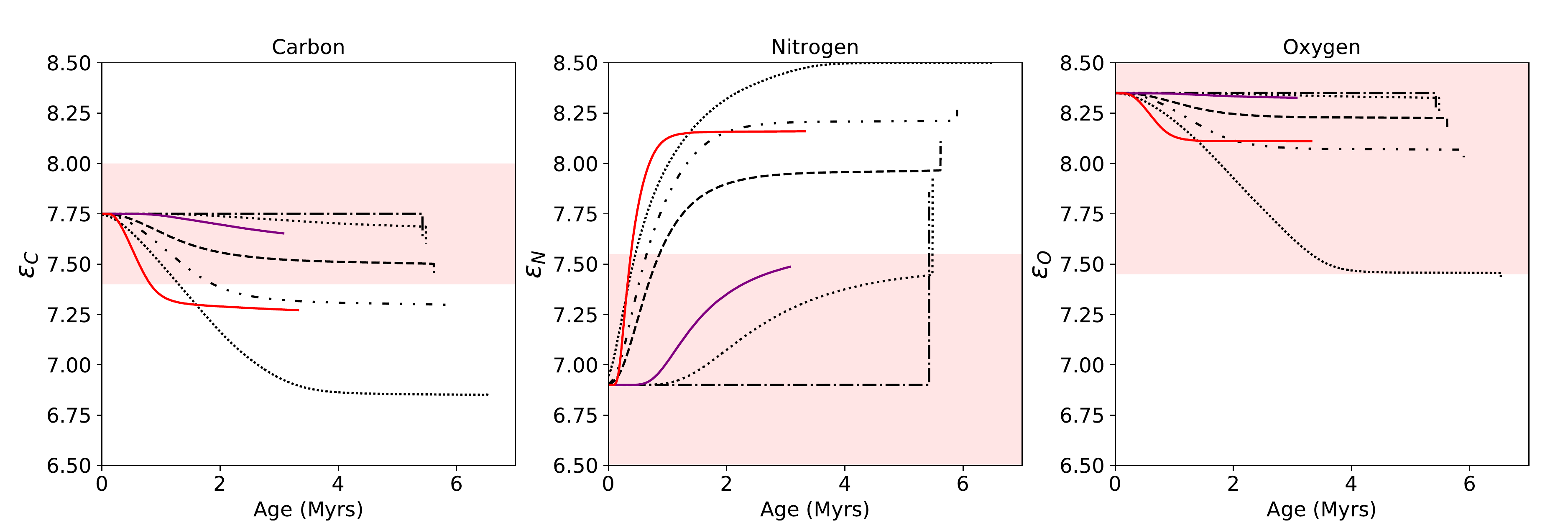}\\
   \includegraphics[width=1\linewidth]{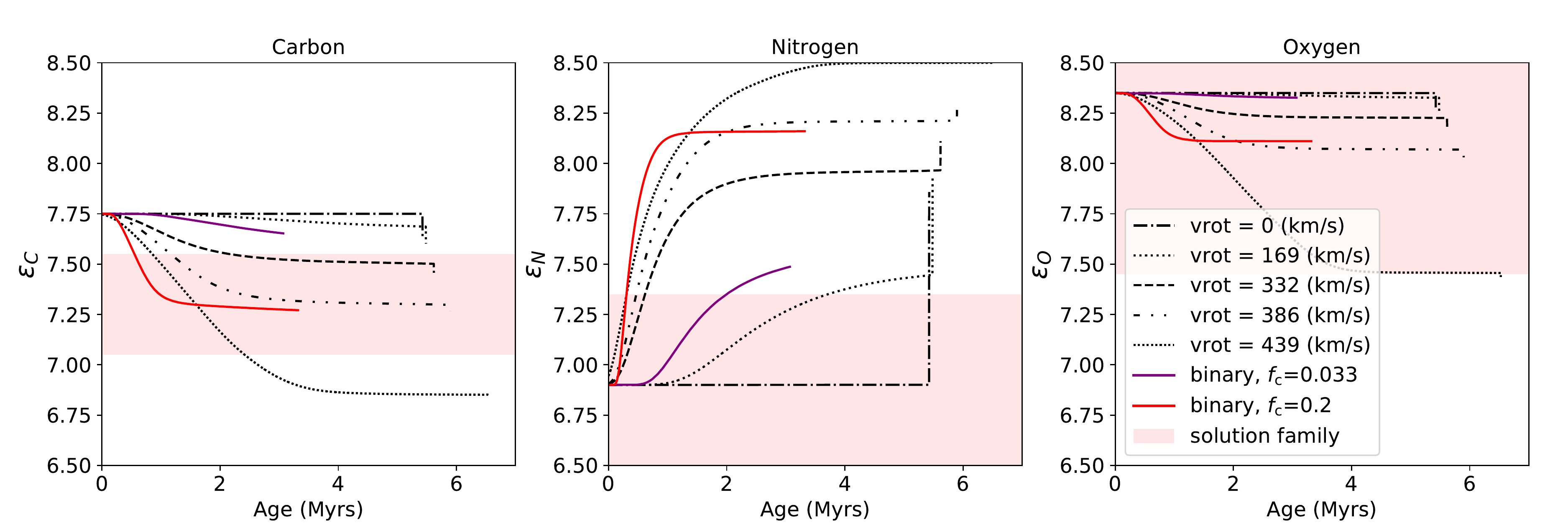}
      \caption{Abundances of carbon (left), nitrogen (middle), and oxygen (right), for the primary (top) and secondary star (bottom) are plotted as a function of age. Single star models from \citet{Brott2011a} (black lines) for initial masses $30\,\mathrm{M}_{\odot}$ and various initial rotational velocities, as our best fit binary model (purple) and our enhanced-mixing binary model (red), are overplotted.  The red shaded regions are the inferred range of abundances produced by our genetic algorithm.
              }
         \label{abundances}
   \end{figure*}

	The same comparison with single-star evolutionary models as the one we did above for \teff\ can be done with surface abundances. Figure \ref{abundances} compares the confidence intervals that we obtained  with the \citet{Brott2011a} 30~$\mathrm{M}_\odot$- evolutionary models with varying rotation rates showing the abundances of CNO elements as a function of  age and range of rotation rates. It also displays our two binary evolutionary models discussed above. With the CNO abundances in agreement with baseline, the primary is either very young (0.5 Myr at most) or does not follow the N-enrichment/C-depletion scheme expected from the rotation rate and high effective temperature. While not shown in Fig. 12, the high effective temperature that we derived would suggest a He enrichment of the order of $\epsilon_\mathrm{He}\sim0.19$ i.e., well above our detection limit and in contradiction with our estimates of  $\epsilon_\mathrm{He}\sim0.11^{+0.02}_{-0.03}$. For the secondary however, neither the single stars nor the binary models can simultaneously match all the inferred surface abundances. The secondary shows no N-enrichment but still seems slightly C-depleted. The oxygen abundance that we derive are compatible with all single and binary evolutionary tracks, and hence provide no discriminatory power.  The Si-abundance is not as sensitive to mixing, and given the derived error bars, we cannot constrain the evolutionary status with Si. The derived nitrogen and helium abundances are compatible with each other for the both the primary and secondary

    The enhanced-mixing binary model comes closer to matching the surface carbon abundance for the secondary, but predicts a higher nitrogen surface abundance, which is not seen. The binary model with a standard mixing parameter has roughly the opposite problem in reproducing the observations. The surface nitrogen abundances are in better agreement while the carbon abundance is more discrepant.

    We conclude that even chemically homogeneously evolving models can not provide a good fit to the VFTS 352 system. They reproduce the HRD position of the stars, but also predict a higher N and He enrichment than currently observed.

	The fundamental limits imposed by the nuclear physics of the CNO cycle mean that reproducing the inferred carbon depletion at the surface of the secondary, whilst avoiding helium and nitrogen surface enhancement seems challenging, impossible even. A possible explanation is that the CNO abundances that we derive are somehow biased, possibly as the results of degeneracies with several wind parameters as discussed above ($\beta$, clumping, etc). As shown in Sect.~\ref{sect: results}, including clumping or a stronger mass loss rate would only worsen the disagreement. Data covering a wider optical range would be useful to access diagnostics lines that should be less affected by the wind properties. Further investigation of the abundances of this system could lift the aforementioned degeneracy between the abundances and the wind parameters.
	

\section{Conclusions} \label{sect: ccl}
    We performed a full atmospheric analysis of the two components of VFTS~352, a massive LMC overcontact binary with an orbital period as short as 1.1~day. Using new phase-resolved HST observations and existing optical spectroscopy, we disentangled  UV-optical spectra of both components and adjusted \textsc{fastwind} NLTE atmosphere models to constrain 11 atmospheric parameters for either stars, including their effective temperature and surface abundances.
    
    The two stars of their respective masses appear hotter than the ZAMS temperature and more luminous than expected for single stars at LMC metallicity. The binary models we computed can reproduce some of the observed orbital and stellar parameters, however they do not reproduce the position of the stars on the HR diagram. Furthermore, we detect no sign of strong He enrichment as expected for stars of such high temperatures. Although the more strongly mixed binary model can roughly reproduce the observed C value in the secondary, they indicate a higher value of N and He than observed in the stars. 
    

One possible explanation for the observed discrepancies is binary interactions.  If the system formed with two unequal mass stars that then underwent a mass transfer event prior to the contact phase, we would expect a measured age discrepancy between the two stars.  This could cause the temperature difference observed between the primary and secondary and may also explain the lack of abundance enhancements.  This would imply that the increased temperature is not due only to mixing and that either the system is too young for mixing to have significantly affected the derived surface abundances or that the mixing behaves in a different way from what theory predicts and suggests.
	
	The anomalous abundances raise many questions. Despite double-checking our results with both the  \textsc{cmfgen} and \textsc{P}o\textsc{WR} NLTE atmospheric models, complementary optical data are highly desirable. 
	The inclusion of \heb~\l4686 and \halpha\ would help to lift some of the possible degeneracies with wind parameters.  Additionally, a broader optical wavelength coverage including several additional carbon and nitrogen lines would help to confirm  the derived abundances for this system.
	
	Putting these findings in the context of chemically homogeneous evolution, VFTS 352 still appears to be a good candidate for future CHE.  The low metallicity, high masses and rotation rates indeed place this object in the regime where CHE is expected to occur \citep{deMink2008, deMink2009}.  We cannot definitively rule out CHE based on our analysis. The system can be young and may not have had enough time to sufficiently mix, however its location in the outskirts of the Tarantula nebula would raise further questions. Alternatively, if the system is older, then the implication is that our understanding of the mixing processes in massive stars may be flawed and may need to be revisited.  Either way, our preliminary binary modeling of this system paves the way for a more extensive investigation in a dedicated paper. 

\begin{acknowledgements}
This work is based on data obtained at the European Southern Observatory under program IDs. 182.D-0222, 090.D-0323, and 092.D-0136.
Support for Program number GO 13806 was provided by NASA through a grant from the Space Telescope Science Institute, which is operated by the Association of Universities for Research in Astronomy, Incorporated, under NASA contract NAS5-26555.
We acknowledge support from the FWO-Odysseus program under project G0F8H6N
This project has received funding from the European Research Council under European Union's Horizon 2020 research programme (grant agreement No 772225)
LAA thanks to Aperfei\c coamento de Pessoal de N\'ivel Superior (CAPES) and Funda\c c\~ao de Amparo \`a Pesquisa do Estado de S\~ao Paulo (FAPESP -- 2011/51680-6,  2012/09716-6, 2013/18245-0) for financial support.
The authors are grateful to Dr. Alex Fullerton, Dr. Elena Sabbi, and Dr. Andrew Tkachenko for useful discussion. 
The authors would also like to thank the referee for their helpful comments and useful discussion.
\end{acknowledgements}




\bibliographystyle{aasjournal}
\bibliography{vfts_bib}

\appendix

\setcounter{figure}{0} \renewcommand{\thefigure}{A.\arabic{figure}}
\section{Parameter Set Descriptions} \label{sect: atm-param-app}
  A brief description of each of our parameters and the lines sensitive to them are listed below.  
	\begin{itemize}
	\item \textbf{Effective temperature (\boldmath$T_\mathrm{eff}$)} - The effective temperature affects the profiles of many lines. In the temperature regime of VFTS 352 ($\gtrsim$ 40\,000K) the lines most affected in the observed optical wavelength range are the helium and hydrogen lines.  In this temperature regime, slight changes in temperature have strong effects on the ionization balance of helium.  Thus, the ratio of He \textsc{i} to He \textsc{ii} (e.g. He \textsc{i} \l4471 : He \textsc{ii} \l4541) is a strong temperature diagnostic; higher temperatures cause stronger He \textsc{ii} lines and weaker He \textsc{i} lines.  The ratio of N \textsc{iv} and N \textsc{v} is also a valid temperature diagnostic in this temperature regime, however the nitrogen lines in the optical portion of our data set are too weak for accurate estimates.  The cores of the Balmer lines are also sensitive to temperature.
		
	\item \textbf{Surface gravity (\boldmath$\log g$)} - The surface gravity affects the wings of the Balmer lines.  A higher surface gravity will make the wings broader while a lower gravity will make the wings narrower. 
			
	\item \textbf{Projected equatorial rotational velocity (\boldmath$v_\mathrm{eq}$ \boldmath$\sin i$)} - The rotational velocity affects all of the lines in the sample.  Higher rotational velocities will broaden lines while low rotational velocities will cause the lines to be narrower.
			
	\item \textbf{Mass loss rate (\boldmath$\log \dot{M}$)} - The mass-loss rate is a wind parameter representing the mass lost in a radially supersonic wind outflow.  Mass lost from Roche lobe overflow or other mechanisms are not considered.  The mass loss rate most strongly affects the shape of C \textsc{iv} \l1548, O \textsc{iv} \l1340 and N \textsc{v} \l1243 in the UV, and of H$\alpha$ and \heb~\l4686 in the optical.  Both H$\alpha$ and \heb~\l4686  are outside the TMBM wavelength range while N \textsc{v}~\l1243 is deeply embed in the wing of Lyman alpha and is thus unusable for VFTS~352.  Aside from the dominant UV-diagnostic lines,  Balmer  and helium lines in the TMBM wavelength range are also affected by the mass loss rate.  As the mass loss rate increases, the emission from the winds fills in the absorption profiles, eventually turning them into emission lines.  

	\item \textbf{Beta (\boldmath$\beta$)} - Beta is a wind parameter that describes the shape of the wind velocity profile.  The velocity of the wind as a function of radial distance from the surface of the star is given by a beta law, the exponent of which is our parameter:
	\begin{equation}
      v(r) = v_\infty \left(1-\frac{bR_*}{r}\right)^\beta,
   	\end{equation}
	
	  where $b$ is obtained from the assumed velocity at the wind-photosphere boundary.  All wind lines are affected by beta but some more than others.  In our data set, the lines that are affected by beta most strongly are C \textsc{iv} \l1548 and O \textsc{iv} \l1340.
			
	\item \textbf{Terminal wind speed (\boldmath$v_\infty$)} - The terminal wind speed defines the maximum speed of the wind.  The terminal wind sets the wavelength of the blue absorption edge of saturated P-Cygni lines and affects the shape of the blue edge of unsaturated P-Cygni lines.  Higher terminal wind speeds in unsaturated lines will cause the slope to be shallower while lower speeds will cause the slope to be sharper.  In our sample, the only P-Cygni line is C \textsc{iv}~\l1548 and it is unsaturated.
			
	\item \textbf{Surface abundances (Helium, Carbon, Nitrogen, Oxygen, Silicon)} - The five abundance parameters behave in the same manner: to strengthen the lines with higher abundances or weaken the lines with lower abundances.  In practice, the abundance of each element only affects lines of that element, with the exception of He that is anti-correlated with that of H.  In the main text, the abundance of helium is given by:
	\begin{equation}
      \varepsilon_\mathrm{He} = \frac{N_\mathrm{He}}{N_\mathrm{H}},
   	\end{equation}	
	
	 where $N_\textrm{He}$ and $N_\textrm{H}$ are the number densities of helium and hydrogen respectively.  The abundance of the other elements is given by:
	  
	\begin{equation}
      \varepsilon_\mathrm{X} = \log \frac{N_\mathrm{X}}{N_\mathrm{H}} + 12,
   	\end{equation}
   	where $N_\textrm{X}$ and $N_\textrm{H}$ are the number densities of the given element and hydrogen respectively.
	\end{itemize}
	
	There are several other parameters that can affect the lines in our sample, however we do not fit over these.  The most important of these parameters and their effects on the shapes of the lines are described below.  To explore how some of these parameters affect our computations, we conducted additional fitting runs where we keep these parameters fixed, however we change the fixed value from that of the original GA run.  These are indicated in the descriptions below.
	
	\begin{itemize}
	\item \textbf{Microturbulent velocity (\boldmath$v_\mathrm{turb}$)} - A microturbulent velocity is considered in the calculation of the atmosphere and wind structure as well as in the formal solution in \textsc{fastwind}.  The microturbulence is set to 15~\kms\ for the calculation of the atmosphere and wind structure but a depth-dependent prescription is used for the calculation of the formal solution as described in Eq.~\ref{microturb_eq} (in this case $v^\mathrm{ph}_\mathrm{turb}$ = 15 km $\mathrm{s}^{-1}$).  The microturbulence affects the shape of all lines and especially the wind lines.  It primarily affects the cores of the lines, changing the line depth, but has a minor impact on the line wings.  For optically thick lines, microturbulence can change the equivalent width, hence affecting the derived abundances.
			
	\item \textbf{Macroturbulent velocity} - Macroturbulence in principle, may affect both the line cores and wings but leaves the equivalent widths unaffected. The stellar components of VFTS~352 are however both rapidly rotating, so that the line broadening is dominated by rotation broadening. In consequence, we neglect the macroturbulence as its contribution is negligible.
	
	\item \textbf{Clumping} - In \textsc{fastwind}, optically thin clumping is parameterized via a clumping factor 
	
	\begin{equation}
	f_\mathrm{cl} = \frac{<\rho^2>}{<\rho>^2} \ge 1, 
	\end{equation}	
	
 which essentially is the multiplication factor that describes how much higher the density within the clump is compared to the mean density. Thus, $f_\mathrm{cl}=1$ implies a smooth uniform wind while $f_\mathrm{cl} > 1$ implies a non-uniform wind with localized regions of high density.  Wind clumping affects different spectral diagnostics differently. As long as clumps are optically thin, diagnostics depending on the square of the density, such as H$\alpha$ in hot stars, become stronger in a clumped wind than in a smooth one of the same mass-loss rate. 

	On the other hand,  diagnostics depending only linearly on density (e.g., UV wind resonance lines) are not directly affected by such optically thin clumping (though indirectly they can be affected through a modified wind ionization balance). Finally, if clumps start to become optically thick, this then leads to an additional leakage of light through porous channels (in physical and/or velocity space) in the wind. For a detailed discussion on  the effect of clumping on the adopted diagnostics, see \citet{Sundqvist2018}.  For our calculations, the clumping factor is set to 1.0, which corresponds to a homogeneous outflow.  We conducted an additional fitting run with the clumping factor set to 10, assuming optically thin clumping. 
	
	\item \textbf{X-rays} - The inclusion of X-rays in the \textsc{fastwind} calculations \citep{Carneiro2016} affects the ionization balance, driving species to higher ionization states.  This then affects the ratios between lines of different ionization states for the same element. For our calculations, we currently do not include X-Rays, however we conduct an additional GA run including an X-Ray radiation field to explore this effect further.
	\end{itemize}

\section{GA results}

   \begin{sidewaysfigure*}
   \centering
   \includegraphics[width=0.9\linewidth]{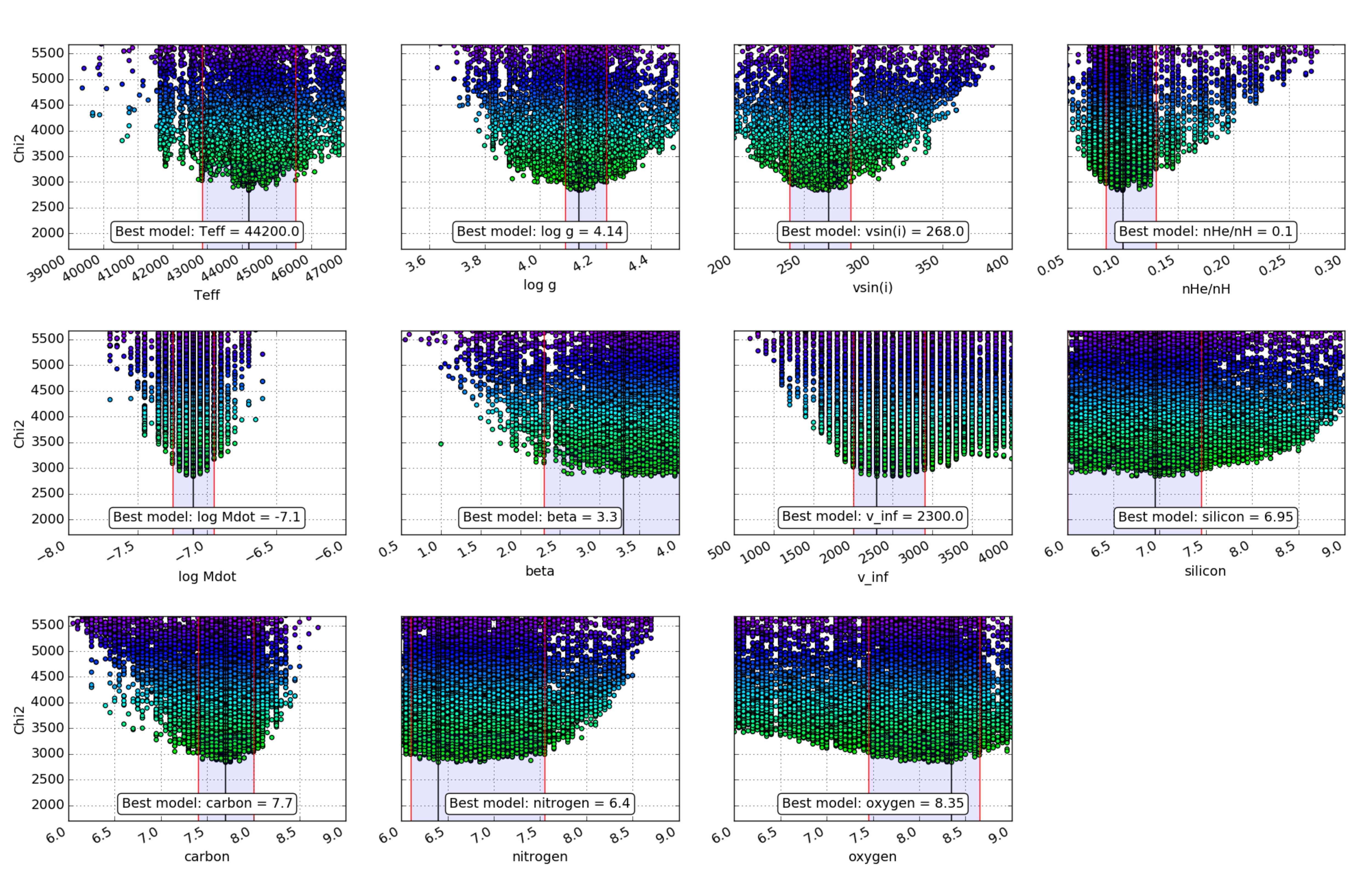}
      \caption{\emph{Primary component:} 11 dimension $\chi^2$\- merit surface projected along the individual  parameter axis. All the $\sim$40\,000 computed \textsc{fastwind} models are included.  The family of acceptable  solutions is indicated with shaded areas while the model with the best $\chi^2$ is indicated by a vertical (black) line within the shaded areas. It is also given on bottom of each panel. 
              }
         \label{ga_fitness_plots-vfts352a}
   \end{sidewaysfigure*}

   \begin{sidewaysfigure*}
   \centering
   \includegraphics[width=0.9\linewidth]{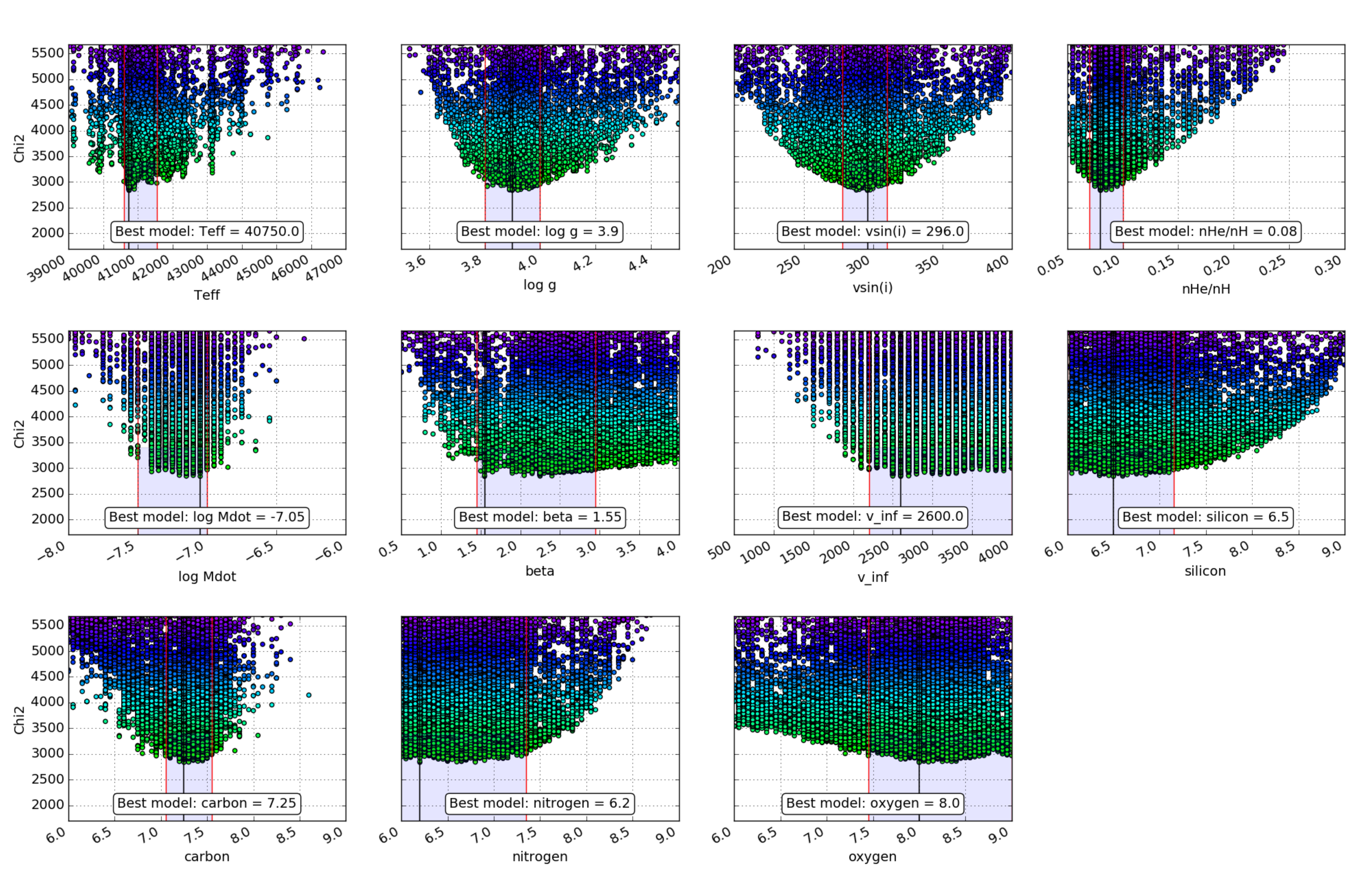}
      \caption{\emph{Secondary component:}  Same as Fig.~\ref{ga_fitness_plots-vfts352a} but for VFTS~352b.
              }
         \label{ga_fitness_plots-vfts352b}
   \end{sidewaysfigure*}




\end{document}